\documentclass[12pt,pdf,epsfig,color,png]{article}
\usepackage{amsmath}
\usepackage{amsfonts}
\usepackage{amsmath}
\usepackage{slashed}
\usepackage{amsxtra}
\usepackage{amstext}
\usepackage{amssymb}
\usepackage{color}
\usepackage{float}
\usepackage{graphicx}
\usepackage{subcaption}
\numberwithin{equation}{section}
\usepackage[dvips]{epsfig}
\usepackage{appendix}
\usepackage{youngtab}
\allowdisplaybreaks

\newcommand{\be}{\begin{equation}}
\newcommand{\ee}{\end{equation}}
\newcommand{\beq}{\begin{equation}}
\newcommand{\eeq}{\end{equation}}
\newcommand{\ba}{\begin{eqnarray}}
\newcommand{\ea}{\end{eqnarray}}
\newcommand{\bea}{\begin{eqnarray}}
\newcommand{\eea}{\end{eqnarray}}

\begin{document}
\baselineskip=15.5pt \pagestyle{plain} \setcounter{page}{1}
%
\begin{titlepage}

\vskip 0.8cm

\begin{center}
%
%
%
%
%

{\Large \bf Fermions hard scattering from superstrings}

\vskip 1.cm

{\large {{\bf Lucas Martin{\footnote{\tt lucasmartinar@iflp.unlp.edu.ar}}, Martin Parlanti{\footnote{\tt martin.parlanti@fisica.unlp.edu.ar}}} {\bf and  Martin
Schvellinger}{\footnote{\tt martin@fisica.unlp.edu.ar}}}}

\vskip 1.cm

{\it Instituto de F\'{\i}sica La Plata-UNLP-CONICET. \\
Boulevard 113 e 63 y 64, (1900) La Plata, Buenos Aires, Argentina \\
and \\
Departamento de F\'{\i}sica, Facultad de Ciencias Exactas,
Universidad Nacional de La Plata. \\
Calle 49 y 115, C.C. 67, (1900) La Plata, Buenos Aires, Argentina.}

\vspace{1.cm}

{\bf Abstract}

\end{center}

\vspace{.5cm}

We study high-energy scattering at fixed angle in the planar limit of non-Abelian gauge field theories from type IIB superstring theory scattering amplitudes. Firstly, we consider four-glueball scattering described in terms of the four-dilaton string theory scattering amplitude. We explicitly calculate some angular integrals of four scalar spherical harmonics on the five sphere leading to selection rules for several scattering processes. Then, we consider the high-energy limit of the four-dilatino scattering amplitude in type IIB superstring theory, which allows to describe four spin-1/2 fermions scattering in the dual gauge field theory. In both cases the partonic structure of the scattering cross sections is obtained. The Regge behavior is also obtained. In addition, we study certain related aspects by considering the effective action constructed from IIB supergravity.

\noindent

\end{titlepage}

\newpage

{\small \tableofcontents}

\newpage

%
%
\section{Introduction}
%
%

We study the high-energy limit at fixed angle of four-point scattering amplitudes of confining non-Abelian gauge field theories in the planar limit in terms of type IIB superstring theory scattering amplitudes. In particular, we consider an IR deformation of the large $N_c$ limit of the strongly-coupled $SU(N_c)$ ${\cal {N}}=4$ SYM theory. Some of the constructions developed in this work are inspired in the hard scattering of an arbitrary number of glueballs represented in terms of the high-energy limit of tree-level closed string scattering amplitudes proposed in reference \cite{Polchinski:2001tt}. The idea of \cite{Polchinski:2001tt} is to consider a local approximation in AdS$_5 \times S^5$ in order to obtain the high-energy limit of the scattering amplitude of glueballs of a four-dimensional dual gauge field theory. For that purpose they carry out an integration of the flat ten-dimensional superstring scattering amplitude on the five sphere and the radial coordinate, using the asymptotic behavior of the wave functions of the external states in AdS$_5 \times S^5$. We extend and further develop this idea to several very interesting directions.

In section 2 we firstly revisit the high-energy limit of two-to-two glueball scattering, including the full calculation of the integrals involving four scalar spherical harmonics on the five sphere, which turns out to be important since it leads to selection rules between the possible asymptotic states in the scattering processes. It is worth to emphasize that we have developed the full type II superstring theory scattering amplitude calculation, including the kinematic factors, the numerical constants as well as the angular integrals of four scalar spherical harmonics on the five sphere which have not been included in previous calculations. The partonic behavior of the total elastic cross section of the two-to-two glueballs scattering is obtained, i.e. $\sigma_{\text{total}}^{\text{elastic}} \propto s^{1-(\Delta_1+\Delta_2)}$, where $s$ is the four-dimensional Mandelstam variable, being $\Delta=k + 4$ (with $k \geq 0$) the conformal dimension of the corresponding spin-zero glueball operator ${\cal{O}}_k^{(8)}(x)$, which in the present case is given by linear combinations of ${\cal {N}}=4$ SYM operators of the form $\text{Tr}(F_+^2 X^k)$ transforming in the $(0, k, 0)$ irreducible representation of the $SU(4)_R$ $R$-symmetry group \cite{DHoker:2002nbb}. The trace is taken on the adjoint representation of $SU(N_c)$. In addition, $F_+$ denotes the self-dual part of the non-Abelian field strength $F=F_+ + F_-$, while  $X$ denotes the 6 real scalar fields, all corresponding to the gauge supermultiplet of ${\cal {N}}=4$ SYM theory in four dimensions. For the total exclusive cross section we obtain $\sigma_{\text{total}}^{\text{exclusive}} \propto s^{1-\Delta/2}$, where $\Delta=\sum_{i=1}^4 \Delta_i$. We also obtain the Regge behavior of the four-dimensional amplitude. Furthermore, we study some important related aspects from the effective five-dimensional supergravity action (after compactificacion on $S^5$) that we present in section 4.

Then, in section 3 we focus on type IIB superstring theory calculations for two-to-two spin-1/2 fermions scattering. This is the main topic of the present work. We obtain original results for fermions which are very interesting for several reasons. The four-dilatino scattering amplitude obtained from type II superstring theory in our work \cite{MPS} allows us to calculate the partonic behavior of the scattering amplitude of the four-dimensional dual gauge theory. Thus, we integrate this ten-dimensional Minkowski space-time scattering amplitude on the radial coordinate and the five sphere considering a local approximation. We obtain an effective four-dimensional scattering amplitude of four spin-$1/2$ fermions in the dual confining four-dimensional gauge field theory. The result gives a high-energy dominant contribution of the form $s^{2-\tau/2}$ plus a sub-leading contribution given by $s^{1-\tau/2}$, where $\tau=\sum_{j=1}^{4}\tau_j$, being $\tau_j$ the twist of the dual gauge theory operator which creates the $j$-th scattered state. For the explicit calculations on the type IIB superstring theory side we restrict to the massless sector, i.e. the type IIB supergravity right-handed dilatinos. Since the number of terms in the scattering amplitude is very large we only present explicitly the calculation of a few representative terms. In this case we consider the operators ${\cal{O}}_k^{(6)}(x)$, which are linear combinations of operators of the form $\text{Tr}(F_+ \lambda_{{\cal {N}}=4} X^k)$, where $\lambda_{{\cal {N}}=4}$ denotes the gauginos of the ${\cal {N}}=4$ SYM theory gauge supermultiplet. The twist is $\tau=\Delta-1/2$, since the spin of these operators is $1/2$, with $\Delta=k+7/2$ and $k \geq 0$. These operators transform in the $(1, k, 0)$ irreducible representation of the $SU(4)_R$ $R$-symmetry group. For instance, for $k=0$ the twist is 3, and these operators transform in the ${\bf {4}}^*$ irreducible representation of $SU(4)_R$. Thus, in this case there are four different spinor spherical harmonics. Each five-dimensional spinor spherical harmonic has four components which implies that for each term in the four-point scattering amplitude we need to solve $4 \times 4 \times 4 \times 4=256$ five-dimensional angular integrals for each of the 4 spinor spherical harmonics in the ${\bf {4}}^*$ irrep. For twist 4 the ${\cal{O}}_1^{(6)}(x)$ operators transform in the ${\bf {20}}^*$ irrep, while for twist 5 the ${\cal{O}}_2^{(6)}(x)$ operators transform in the ${\bf {60}}^*$ irrep, just to give a few examples, being the number of angular integrals very large. Although there are explicit expressions for the spinor spherical harmonics on the five sphere, due to the described complexity of the full calculation we have not calculated explicitly their angular integrals. For the elastic scattering process of four spin-1/2 fermions we obtain the total cross section $\sigma_{\text{total}}^{\text{elastic}} \propto s^{1-(\tau_1+\tau_2)}$, while the total exclusive cross section is $\sigma_{\text{total}}^{\text{exclusive}} \propto s^{1-\tau/2}$, where $\tau=\sum_{j=1}^4 \tau_j$. On the other hand, we calculate both the fixed angle and the Regge limit. We also comment on interesting aspects related to the effective five-dimensional action for spin-1/2 fermions including the Pauli terms. These ideas are discussed in section 4.

In the case of two-to-two glueballs scattering the effective action written in terms of type IIB supergravity fields leads to a $t$-channel with the exchange of a graviton propagating within the AdS$_5$ space. This is consistent with the Regge limit of the four-dilaton superstring scattering amplitude, being $j=2$ the leading power of the Mandelstam variable $\tilde{s}$, corresponding to the exchange of a single Reggeized graviton. 
On the other hand, the high-energy limit of four spin-1/2 fermions scattering amplitude contains two important contributions. The dominant one corresponds to the exchange of a single Reggeized graviton, leading to $j=2$. There is also a sub-dominant contribution which involves the exchange of a single Reggeized $B^1_\mu$ field (with $j=1$). This vector field $B^1_\mu$ is a linear combination of the gravi-photon $h_{\mu \alpha}$ and the Ramond-Ramond four-form field $A_{\mu \alpha \beta \delta}$ with the Lorentz index $\mu$ on the AdS$_5$, while indices $\alpha$, $\beta$, $\delta$ correspond to $S^5$. The dilatino part of the five-dimensional effective action was derived from type IIB supergravity compactified on $S^5$ in \cite{Jorrin:2020cil,Jorrin:2020kzq}, and it includes Pauli interaction terms which play an essential role in the exchange of the vector field.

The partonic behavior in terms of the hard-wall and the soft wall \cite{Karch:2006pv} models as well as the Witten-Sakai-Sugimoto model \cite{Witten:1998qj,Sakai:2004cn} has been investigated in reference \cite{Bianchi:2021sug}. They have generalized the proposal described in \cite{Polchinski:2001tt} to the mentioned confining backgrounds for the case of glueball scattering amplitudes in the dual gauge field theory. They also have considered the expansion of the scattering amplitudes around their poles, integrated in the radial coordinate, and carried out a re-summation of the resulting expansion. In addition, they have calculated the open-string scattering amplitudes and obtained the corresponding ones for meson scattering. On the other hand, in references \cite{Domokos:2009hm,Domokos:2010ma} it has been investigated the Pomeron contribution to proton-proton and proton-anti proton scattering, in terms of the Sakai-Sugimoto model and the AdS/QCD model, respectively. In \cite{Domokos:2009hm} it has been shown a very good level of agreement with experimental data for the total and differential cross-sections. In addition, deep inelastic scattering (DIS) has been studied in terms of the gauge/string theory duality in the context of top-down models after the pioneering work of Polchinski and Strassler \cite{Polchinski:2002jw}. They have considered glueballs in the parametric regimes of the Bjorken parameter where either supergravity or string theory become dominant, and also the IIB supergravity regime for DIS from dilatinos. Extensions of this idea in the context of type IIB superstring theory for spin-1/2 fermions have been considered in \cite{Kovensky:2018xxa}. Detailed investigations in terms of type IIB supergravity for twist 3 and higher twist operators have been done in \cite{Jorrin:2020cil} and \cite{Jorrin:2020kzq}, respectively. Top-down holographic dual descriptions of DIS off mesons have been considered in references \cite{Koile:2011aa,Koile:2013hba,Jorrin:2016rbx,Jorrin:2016ccr}, obtaining a reasonably good comparison for the first three moments of the $F_2$ structure function of the pion and the first three moments of the $F_1$ structure function of the $\rho$ meson with respect to the lattice QCD calculations. For baryons a crucial development was the so-called Brower-Polchinski-Strassler-Tan (BPST) Pomeron \cite{Brower:2006ea}, which leads to a unified description of the soft Pomeron and the BFKL Pomeron. The BPST Pomeron gives an excellent agreement with experimental data for the structure function $F_2$ \cite{Brower:2010wf,Jorrin:2022lua} (it fits 280 experimental data within the Bjorken parameter range $x_B \leq 0.01$ and the squared virtual-photon momentum range $0.1$ GeV$^2 \leq Q^2 \leq 400$ GeV$^2$, using only four free parameters and $\chi^2_{\text{d.o.f.}}=1.086$ \cite{Jorrin:2022lua}), while the so-called Holographic $A_4$ Pomeron \cite{Kovensky:2018xxa,Jorrin:2022lua} gives an excellent fit to experimental data for the function $g_1$ related to the spin structure of the proton (it fits 55 experimental data with only one free parameter and $\chi^2_{\text{d.o.f.}}=1.14$ \cite{Jorrin:2022lua}, in this case for $x_B \leq 0.01$ and  $0.062$ GeV$^2 \leq Q^2 \leq 2.41$ GeV$^2$), and also it leads to sharp predictions for $g_1$ for the Electron-Ion Collider at very small values of the Bjorken parameter \cite{Borsa:2023tqr}. The proton structure functions $F_2$ and $x F_3$ have also been studied in terms of the BPST and the Holographic $A_4$ Pomerons, respectively, in the Bjorken parameter range $[0.01, 0.1]$ and for much larger values of $Q^2$ where neutral-current electroweak effects become dominant. These results are in an excellent agreement with experimental data \cite{Chaves}.

%
%
\section{High-energy limit of four glueballs scattering amplitude}
%
%

In this section we revisit in detail the calculation carried out in \cite{Polchinski:2001tt}. This is very instructive in order to be able to compare these results with the ones from fermion scattering amplitudes from type IIB superstring theory presented in section 3. We also explicitly solve the angular integrals on the five sphere which were not considered in \cite{Polchinski:2001tt} and previous papers where a general compact manifold was studied\footnote{We consider the specific case of the five sphere in order to be able to obtain specific selection rules for an IR deformation $SU(N_c)$ ${\cal {N}}=4$ SYM theory. The methods discussed in the present work in principle can be extended to other compact manifolds.}. This allows us to find selection rules establishing which states can or cannot be involved in these scattering processes. Moreover, in section 4 we will further discuss on this point in relation to the effective five-dimensional action obtained from type IIB supergravity.

Polchinski and Strassler \cite{Polchinski:2001tt} investigated the high-energy limit of $2 \rightarrow m$ glueballs in confining gauge theories at fixed angle in the framework of the gauge/string theory duality. They concluded that when strings propagate in certain curved backgrounds the corresponding scattering amplitudes show a power-law behavior as a function of the center-of-mass energy $\sqrt{s}$. Moreover, they found hard-scattering behavior in a certain kinematic range and Regge behavior for a different range.

In order to have a mass gap in the four-dimensional confining gauge theory, one can set a scale $\Lambda$ of the order of the lightest glueball mass. It corresponds to a lower cut-off in the radial coordinate on the dual string theory background, namely: $r_0 \sim \Lambda R^2 \leq r$, being $R^4 = 4 \pi g_s N_c \alpha'^2$, where $g_s$ is the string coupling while $\alpha'$ is the string constant. We work in natural units $c=\hbar=1$. The metric of AdS$_5 \times S^5$ space is given by
\begin{eqnarray}
ds^2 = \frac{r^2}{R^2} \eta_{\mu \nu} dx^\mu dx^\nu + \frac{R^2}{r^2} dr^2 + R^2 d\Omega_5^2 \ .
\end{eqnarray}
The four-momentum of the dual gauge theory is $p_\mu=-i \partial_\mu$. Given an inertial observer localized at position $r$ in the AdS bulk, its ten-dimensional momentum is given by
\begin{eqnarray}
\tilde{p}^\mu = \frac{R}{r} p^\mu \ , \label{losp}
\end{eqnarray}
which is a crucial point since it signals that the energy scale for a given process in the gauge theory is related to a certain dual string theory scattering process at the position $r$ in the AdS$_5$ space. The string tension in the gauge theory is $\hat{\alpha}' \equiv 1/(\lambda_{\text{'t Hooft}}^{1/2} \Lambda^2)$, with the 't Hooft coupling defined as $\lambda_{\text{'t Hooft}} = g_{YM}^2 N_c \equiv 4 \pi g_s N_c$. 
In addition, there is the following relation
\begin{eqnarray}
\sqrt{\alpha^\prime} \tilde{p}= \sqrt{\hat{\alpha}^\prime} p \frac{r_0}{r} 
\, .
\end{eqnarray}

In this section we are interested in the two-to-two scattering of glueball states associated with $SU(N_c)$ ${\cal {N}}=4$ SYM theory with an IR deformation inducing confinement. The glueball states of the gauge theory correspond to closed string states. In particular, they are related to the dilaton in the gauge/string theory duality. We consider the dilaton wave-function given by
\begin{eqnarray}
\Phi(x, r, \Omega) = e^{i p \cdot x} \psi(r, \Omega) \, . \label{dilaton-10d}
\end{eqnarray}
Since the variation of the exponential depends on $r$ through $p \sim r/R^2$, the wave function $e^{i p \cdot x}$ varies on the string scale. On the other hand, since the gauge/string theory duality in this case holds for $N_c \gg \lambda_{\text{'t Hooft}}^{1/2} \gg 1$, the variation of the function $\psi(r, \Omega)$ is slow. This implies that the closed string scattering effectively occurs at $(r, \Omega)$, and then the effective four-dimensional scattering amplitude can be obtained by the integration of the  scattering amplitude of four closed strings over these coordinates as follows:
\begin{equation}
\mathcal{A}_{4}^{4d}(p) = \int_{r_0}^{\infty} dr \int_{S^5} d\Omega_5 \ \sqrt{-g} \ \mathcal{A}_{\text{string}}^{10d}(\tilde{p}) \, , \label{A4}
\end{equation}
where $g$ is the determinant of the ten-dimensional metric, while the flat-space string theory scattering amplitude $\mathcal{A}_{\text{string}}^{10d}(\tilde{p})$ can be written as\footnote{Recall that we consider a local approximation, for which we start from the flat-space ten-dimensional closed string scattering amplitude. This leads to an effective description based on a coherent integration over all possible localizations in $(r, \Omega)$ of the string theory scattering process in the ten-dimensional bulk.} 
\begin{eqnarray}
\mathcal{A}_{\text{string}}^{10d}(\tilde{p}) = 4 g_s^2 \alpha'^3 F(\tilde{p}\sqrt{\alpha^\prime}) \, ,
\end{eqnarray}
where
\begin{eqnarray}
F(\tilde{p} \sqrt{\alpha'}) = \prod_{\chi=\tilde{s},\tilde{t},\tilde{u}} \frac{\Gamma(-\alpha' \chi/4)}{\Gamma(1+\alpha' \chi/4)} \ K(\tilde{p} \sqrt{\alpha'}) \, . \label{FGammaK} 
\end{eqnarray}
The ten-dimensional Mandelstam variables are defined as follows:
\begin{equation}
\tilde{s}=-(k_1+k_2)^2 \, , \,\,\,\,\,\,\, 
\tilde{t}=-(k_1+k_4)^2 \, , \,\,\,\,\,\,\,
\tilde{u}=-(k_1+k_3)^2 \, , 
\end{equation}
where $k_i$'s denote the four ten-dimensional incoming momenta. Also, notice that 
\begin{equation}
K(\tilde{p}\sqrt{\alpha'})= - \pi \alpha'^4 K_{\text{closed}}(k_1,k_2,k_3,k_4) \, , \label{K2-9}
\end{equation}
being $K_{\text{closed}}(k_1,k_2,k_3,k_4)$ the kinematic factor of four closed strings scattering amplitude. In terms of the Kawai-Lewellen-Tye relation \cite{Kawai:1985xq} this can be written as the tensor product of two kinematic factors of four open strings scattering amplitude as follows
\begin{eqnarray}
K_{\text{closed}}(k_1,k_2,k_3,k_4)= K_{\text{open}}(k_1/2,k_2/2,k_3/2,k_4/2)\otimes \tilde{K}_{\text{open}}(k_1/2,k_2/2,k_3/2,k_4/2) \, , 
&& \label{Kclosed}
\end{eqnarray}
where the kinematic factor of four open strings scattering amplitude is \cite{Green:1987sp}
\begin{eqnarray}
K_{\text{open}}(k_1,k_2,k_3,k_4) &=& -\frac{1}{4} \ (\tilde{s} \tilde{t} \ \zeta_1 \cdot \zeta_3 \ \zeta_2 \cdot \zeta_4
+ \tilde{s} \tilde{u} \ \zeta_2 \cdot \zeta_3 \ \zeta_1 \cdot \zeta_4
+ \tilde{t} \tilde{u} \ \zeta_1 \cdot \zeta_2 \ \zeta_3 \cdot \zeta_4) \nonumber \\
&&
+ \frac{\tilde{s}}{2} \left(\zeta_1 \cdot k_4  \ \zeta_3 \cdot k_2 \ \zeta_2 \cdot \zeta_4 + \zeta_2 \cdot k_3  \ \zeta_4 \cdot k_1 \ \zeta_1 \cdot \zeta_3 \right. \nonumber \\
&& +\left. \zeta_1 \cdot k_3  \ \zeta_4 \cdot k_2 \ \zeta_2 \cdot \zeta_3 + \zeta_2 \cdot k_4  \ \zeta_3 \cdot k_1 \ \zeta_1 \cdot \zeta_4 \right) \nonumber \\
&&
+ \frac{\tilde{t}}{2} \left(\zeta_2 \cdot k_1  \ \zeta_4 \cdot k_3 \ \zeta_3 \cdot \zeta_1 + \zeta_3 \cdot k_4  \ \zeta_1 \cdot k_2 \ \zeta_2 \cdot \zeta_4 \right. \nonumber \\
&& +\left. \zeta_2 \cdot k_4  \ \zeta_1 \cdot k_3 \ \zeta_3 \cdot \zeta_4 + \zeta_3 \cdot k_1  \ \zeta_4 \cdot k_2 \ \zeta_2 \cdot \zeta_1 \right) \nonumber \\
&& + \frac{\tilde{u}}{2} \left(\zeta_1 \cdot k_2  \ \zeta_4 \cdot k_3 \ \zeta_3 \cdot \zeta_2 + \zeta_3 \cdot k_4  \ \zeta_2 \cdot k_1 \ \zeta_1 \cdot \zeta_4 \right. \nonumber \\
&& +\left. \zeta_1 \cdot k_4  \ \zeta_2 \cdot k_3 \ \zeta_3 \cdot \zeta_4 + \zeta_3 \cdot k_2  \ \zeta_4 \cdot k_1 \ \zeta_1 \cdot \zeta_2 \right) \, . \label{Kopen4}
\end{eqnarray}
By using this equation and (\ref{Kclosed}) it generates 225 terms, however there are only 120 independent ones. These terms can be calculated from the tensor product $\zeta_i^M \otimes \zeta_i^{M'}$, where $\zeta_i^M$ is a vector polarization, being $M$ and $M'$ ten-dimensional Lorentz indices. The polarization of massless particles in the closed string can be written as a second rank tensor \cite{Gross:1986mw}
\begin{eqnarray}
\Theta_i^{MM'}=\zeta_i^M \otimes \zeta_i^{M'} \, .
\end{eqnarray}
Then, for the dilaton we have 
\begin{eqnarray}
\Theta_i^{MM'}=\frac{1}{\sqrt{8}} \Phi_i \left(\eta^{MM'}-k_i^M\bar{k}_i^{M'}-k_i^{M'}\bar{k}_i^{M}\right) \, , 
\end{eqnarray}
where for each particle $i$ it has been introduced the auxiliary momentum $\bar{k}_i$ satisfying the relations $k \cdot \bar{k}=1$ and $\bar{k} \cdot \bar{k}=0$. Notice that $\Phi_i$ denotes the dilaton wave function of the $i$-th particle. Since we only consider massless string theory states it follows the kinematic relation $\tilde{s}+\tilde{t}+\tilde{u}=0$, as well as the dispersion relations $k_i \cdot k_i =0$. These properties make easier the lengthy calculation, obtaining
\begin{eqnarray}
K_{\text{closed}}(k_1,k_2,k_3,k_4)&=&\frac{1}{128} [6 \tilde{s}^4-15\tilde{s}^3(\tilde{t}+\tilde{u})+\tilde{s}^2(30 \tilde{t}^2+28\tilde{t}\tilde{u}+30 \tilde{u}^2) \nonumber \\
&-&\tilde{s}(\tilde{t}+\tilde{u})(15 \tilde{t}^2-43\tilde{t}\tilde{u}+15\tilde{u}^2)\nonumber\\
&+& 3(2\tilde{t}^4-5\tilde{t}^3\tilde{u}+10\tilde{t}^2\tilde{u}^2-5\tilde{t}\tilde{u}^3+2 \tilde{u}^4)] \Phi_1 \Phi_2 \Phi_3 \Phi_4 \, .
\end{eqnarray}
Since we consider the dilaton in ten dimensions, we have $\tilde{u}=-\tilde{s}-\tilde{t}$, therefore the previous expression becomes very simple
\begin{eqnarray}
K_{\text{closed}}(k_1,k_2,k_3,k_4)&=& \frac{9}{16}\left(\tilde{s}^2+\tilde{s}\tilde{t}+\tilde{t}^2\right)^2 \Phi_1 \Phi_2 \Phi_3 \Phi_4  \, .
\end{eqnarray}
Another important kinematic properties are the relations between the Mandelstam variables and the scattering angle $\theta$, given by
\begin{equation}
\tilde{t}= -\frac{\tilde{s}}{2} (1-\cos\theta) \, , \,\,\,\,\,\,\,\, \text{and} \,\,\,\,\,\,\,\, \tilde{u}=-\frac{\tilde{s}}{2} (1+\cos\theta) \, . \label{kinematics}
\end{equation}
Thus, we can work out  $K_{\text{closed}}(k_1,k_2,k_3,k_4)$ obtaining\footnote{Notice that the same result is obtained from $K_{\text{closed}}(k_1,k_2,k_3,k_4)=\frac{9}{32} \left(\tilde{s}^4+\tilde{t}^4+\tilde{u}^4\right) \Phi_1 \Phi_2 \Phi_3 \Phi_4$.}
\begin{eqnarray}
K_{\text{closed}}(k_1,k_2,k_3,k_4) &=& \frac{9 \tilde{s}^4 (\cos (2 \theta)+7)^2}{1024}
 \ \Phi _1 \Phi_2 \Phi_3 \Phi_4  \, , \label{K2-15}
\end{eqnarray}
which for the ten-dimensional Minkowski space-time scattering amplitude of four dilatons leads to
\begin{eqnarray}
\mathcal{A}_{\text{string}}^{10d, {\text{4 dilatons}}}(\tilde{p}) &=&-4 \pi g_s^2 \alpha^{\prime 3}\frac{9 (\alpha'\tilde{s})^4 (\cos (2 \theta)+7)^2}{1024} \prod_{\chi=\tilde{s},\tilde{t},\tilde{u}} \frac{\Gamma(-\alpha^\prime \chi/4)}{\Gamma(1+\alpha^\prime \chi/4)} \ \Phi_1 \Phi_2 \Phi_3 \Phi_4 \, , \nonumber \\
\end{eqnarray}
where the ten-dimensional Newton constant is $\kappa_{10}^2=4 g_s^2 \alpha'^4$.

Now, let us focus on the dilaton wave functions $\Phi_i$'s.
Since we are interested in the high-energy behavior of the scattering amplitude, i.e. the UV dynamics of the gauge theory, we consider large $r$ values. Then, from the dilaton wave function (\ref{dilaton-10d}) we may write
\begin{eqnarray}
\psi_{\Delta_i} (r,\Omega) \approx \frac{c_i}{R^4\Lambda} \left(\frac{r_0}{r}\right)^{\Delta_i}  Y_{\Delta_i}(\Omega) \ ,
\end{eqnarray}
where $c_i$ is a normalization constant, $Y_{\Delta_i}(\Omega)$ 
is a scalar spherical harmonic on $S^5$, and $\Delta_i$ is the conformal dimension of the lowest conformal dimension operator which creates the corresponding state. These wave functions satisfy the orthonormalization condition 
\begin{eqnarray}
\int_{r_0}^{\infty} dr  \int_{S^5} d \Omega_5 \ \sqrt{g_{\perp}} \ \frac{r^2}{R^2} \ \psi_{\Delta_i}^*(r,\Omega) \ \psi_{\Delta_j}(r,\Omega) = \delta_{\Delta_i \Delta_j} \, ,
\end{eqnarray}
with $\sqrt{g_{\perp}}=\frac{R^6}{r}$. In addition, the orthonormalization condition for the scalar spherical harmonics on $S^5$ is given by
\begin{equation}
\delta_{\Delta_i,\Delta_j} = \int d \Omega_5 \ \sqrt{\hat{g}_{S^5}} \ Y^*_{\Delta_i}(\Omega) \ Y_{\Delta_j}(\Omega) \  ,
\end{equation}
where the unit five-sphere leads to $\sqrt{\hat{g}_{S^5}}=1$ and $\sqrt{g_{S^5}}=R^5$, while the radial integral gives
\begin{eqnarray}
R^4 \int_{r_0}^\infty dr \ r \ \frac{|c_i|^2}{R^8\Lambda^2}  \left(\frac{r_0}{r}\right)^{2\Delta_i} = 1 \, ,
\end{eqnarray}
which fixes the normalization constant $|c_i| = \sqrt{2(\Delta_i-1)}$. Therefore, the amplitude (\ref{A4}) becomes
\begin{eqnarray}
\mathcal{A}^{4d}_{4 \ {\text{glueballs}}}(p)&=& -144 \pi \ \frac{(\cos(2 \theta) + 7)^2}{1024} \  \frac{g_s^2 \ \alpha'^3}{R^6 r_0^{4-\Delta}} \ I_{\Delta_1,\Delta_2,\Delta_3,\Delta_4}\ \left(\prod_{i=1}^4 (\Delta_i-1)^{1/2}\right) \times \nonumber \\
&& \int_{r_0}^\infty dr \ r^{3-\Delta} \ (\alpha' \tilde{s})^4 \prod_{\chi=\tilde{s},\tilde{t},\tilde{u}} \frac{\Gamma(-\alpha^\prime \chi/4)}{\Gamma(1+\alpha^\prime \chi/4)} \, ,  \label{A4-F}
\end{eqnarray}
where $\Delta=\sum_{i=1}^4 \Delta_i$, while the angular integral of four scalar spherical harmonics on $S^5$ is given by
\begin{equation}
I_{\Delta_1,\Delta_2,\Delta_3,\Delta_4} = \int d \Omega_5 \ \sqrt{\hat{g}_{S^5}} \ Y^*_{\Delta_1}(\Omega) \ Y_{\Delta_2}(\Omega) \ Y^*_{\Delta_3}(\Omega) \ Y_{\Delta_4}(\Omega) \ . \label{Integral-4-scalars}
\end{equation}
Let us firstly consider the radial integral in the amplitude (\ref{A4-F}). Later on we shall return to the angular integral on $S^5$ of equation (\ref{Integral-4-scalars}) which leads to certain selection rules. The amplitude (\ref{A4-F}) is dominated by $r_{\text{scat}} \sim R \sqrt{\alpha'} p \sim r_0 \sqrt{\hat{\alpha'}} p$, where effectively the closed string scattering occurs. In order to study this radial integral let us firstly consider that $\alpha' \tilde{s}/4 \gg 1$ (with $\alpha' \tilde{|t|}$ and $\alpha' \tilde{|u|} \gg 1$), and take into account the kinematical relation $\tilde{s}+\tilde{t}+\tilde{u}=0$. We may use the asymptotic representation of the Gamma function for large values of $|z|$, $\Gamma(z) = \sqrt{2 \pi} \ z^{z-1/2} \ e^{-z} (1+1/(12 z)+\dots)$, which holds for $|\arg z|<\pi$ (see for instance equation (8.327) of \cite{Gradshteyn:1943cpj}). Thus, we have the following approximation,
\begin{eqnarray}
\prod_{\chi=\tilde{s},\tilde{t},\tilde{u}} \frac{\Gamma(-\alpha^\prime \chi/4)}{\Gamma(1+\alpha^\prime \chi/4)} &=&
(-1)^{\alpha'\tilde{s}/2+1} \frac{128 \ e^2}{\alpha'^3}\frac{ \sin(\frac{\pi \alpha'\tilde{t}}{4}) \sin(\frac{\pi \alpha'\tilde{u}}{4})}{\sin(\frac{\pi \alpha'\tilde{s}}{4})} 
\frac{e^{-2\beta_{\tilde{s}\tilde{t}\tilde{u}}}}{\tilde{s}\tilde{t}\tilde{u}} \, , \label{gamma-products}
\end{eqnarray} 
where $e$ is the Euler number, and 
\begin{eqnarray}
\beta_{\tilde{s}\tilde{t}\tilde{u}}=\frac{\alpha'\tilde{s}}{4}\log\left(\frac{\alpha'\tilde{s}}{4}\right)+\frac{\alpha'\tilde{t}}{4} \log\left(\frac{\alpha'\tilde{t}}{4}\right)+\frac{\alpha'\tilde{u}}{4} \log\left(\frac{\alpha'\tilde{u}}{4}\right) \, ,
\end{eqnarray}
which by using the kinematic relations (\ref{kinematics}) becomes
\begin{eqnarray}
\beta_{\tilde{s}\tilde{t}\tilde{u}}=-\frac{\alpha'\tilde{s}}{4}\Bigg( i\pi+\frac{(1-\cos\theta)}{2} \log\left(\frac{1-\cos\theta}{2}\right)+\frac{(1+\cos\theta)}{2} \log\left(\frac{1+\cos\theta}{2}\right)\Bigg) \, . \label{beta}
\end{eqnarray}
Considering the relation between the Mandelstam variables in four and ten dimensions,
\begin{eqnarray}
\tilde{s}=\frac{R^2}{r^2}s \, , 
\end{eqnarray}
the amplitude $\mathcal{A}^{4d}_{4 \ {\text{glueballs}}}(p)$ can be written as
\begin{eqnarray}
\mathcal{A}^{4d}_{4 \ {\text{glueballs}}}(p)&\approx& 72 \pi \ \frac{(\cos(2 \theta) + 7)^2}{\sin^2\theta} \  \frac{g_s^2 \ \alpha'^3 \ e^2}{R^6 r_0^{4-\Delta}} \ I_{\Delta_1,\Delta_2,\Delta_3,\Delta_4}\ \left(\prod_{i=1}^4 (\Delta_i-1)^{1/2}\right) \times \nonumber \\
&& \int_{r_0}^\infty dr \ r^{3-\Delta} \ \left(\alpha' \frac{R^2 s}{r^2} \right)
(-1)^{\frac{\alpha' R^2 s}{2 r^2}} \frac{ \sin(\frac{\pi \alpha'\tilde{t}}{4}) \sin(\frac{\pi \alpha'\tilde{u}}{4})}{\sin(\frac{\pi \alpha'\tilde{s}}{4})} 
e^{-2\beta_{\tilde{s}\tilde{t}\tilde{u}}} \, ,
\end{eqnarray}
where the approximation means the use of the Stirling's formula. Also notice that there is an oscillating function within the integrand, 
\begin{eqnarray}
(-1)^{\frac{\alpha' R^2 s}{2 r^2}} \sin\left(\frac{\pi \alpha' R^2 s (1-\cos\theta)}{8 r^2}\right) \sin\left(\frac{\pi \alpha' R^2 s (1+\cos\theta)}{8 r^2}\right)\sin^{-1}\left(\frac{\pi \alpha' R^2 s}{4 r^2}\right) \, , \label{Imaginary-part-factor}
\end{eqnarray}
which contains the zeros and poles of the scattering amplitude. By integration on the radial coordinate an average behavior at high energy emerges \cite{Bianchi:2021sug}. For a fixed value of the four-dimensional Mandelstam variable and large $r$, this function mainly depends on the scattering angle $\theta$. Moreover, it does not lead to contributions to the exponent of this Mandelstam variable. However, we should emphasize that from this factor it comes an imaginary part of the scattering amplitude (\ref{Glueball-amplitude}), which is related to the total cross section as shown in equation (\ref{Total-cross-section-glueballs}). Thus, we must solve the radial integral as in \cite{Bianchi:2021sug}
\begin{eqnarray}
I_r
&=& \int_{r_0}^\infty dr \ r^{3-\Delta} \ \left(\alpha' \frac{R^2 s}{r^2} \right) \
e^{-2\beta_{\tilde{s}\tilde{t}\tilde{u}}} \, .
\end{eqnarray}
Including the result of this integral we have
\begin{eqnarray}
\mathcal{A}^{4d}_{4 \ {\text{glueballs}}}(p)&\approx& 72 e^2 \pi \ \frac{(\cos(2 \theta) + 7)^2}{\sin^2\theta} \  \frac{g_s^2 \alpha'^3}{R^6} \ I_{\Delta_1,\Delta_2,\Delta_3,\Delta_4}\ \left(\prod_{i=1}^4 (\Delta_i-1)^{1/2}\right) \times \nonumber \\
&& (-1)^{1-\Delta/2} \ 2^{\Delta-3} \left(\frac{\alpha'}{\Lambda^2 R^2}\right)^{2-\Delta/2} f(\theta)^{1-\Delta/2} \ s^{2-\Delta/2} \times  \nonumber \\
&& \left\{\Gamma\left(\frac{\Delta-2}{2}\right)-\Gamma\left(\frac{\Delta-2}{2}, -\frac{\alpha' s}{4 \Lambda^2 R^2} f(\theta) \right)\right\} \, , \label{Glueball-amplitude}
\end{eqnarray}
where the Gamma function as well as the incomplete Gamma function come from the radial integration, and we have defined
\begin{eqnarray}
f(\theta) &=& 2 \pi i + (1+\cos\theta) \log[\cos^2(\theta/2)]+(1-\cos\theta) \log[\sin^2(\theta/2)] \, .
\end{eqnarray}
The incomplete Gamma function can be asymptotically expanded as follows (equation (6.5.32) of \cite{Abramowitz})
\begin{equation}
\Gamma(a, z) \equiv \int_z^\infty dt \, t^{a-1} \, e^{-z} \approx z^{a-1} e^{-z} \left[1+\frac{a-1}{z}+\frac{(a-1)(a-2)}{z^2}+\dots\right] \, ,
\end{equation}
as $z \rightarrow \infty$ and $|\text{arg}(z)|<3 \pi/2$. We have explicitly checked that in the high-energy limit it vanishes
\begin{eqnarray}
\text{limit}_{s\rightarrow\infty} \Gamma\left(\frac{\Delta-2}{2}, -\frac{\alpha' s}{4 \Lambda^2 R^2} f(\theta) \right)
=0 \, .
\end{eqnarray}
Therefore, the high-energy limit of two-to-two glueballs scattering amplitude is
\begin{eqnarray}
\mathcal{A}^{4d}_{4 \ {\text{glueballs}}}(p)&\approx& \frac{9 \ e^2}{2 \pi} \ \frac{(\cos(2 \theta) + 7)^2}{\sin^2\theta} \  \frac{(4 \pi g_s N_c)^{\frac{\Delta-2}{4}}}{N_c^2} \ I_{\Delta_1,\Delta_2,\Delta_3,\Delta_4}\ \left(\prod_{i=1}^4 (\Delta_i-1)^{1/2}\right) \times \nonumber \\
&& (-1)^{1-\Delta/2} \ 2^{\Delta-3} f(\theta)^{1-\Delta/2} \ \Gamma\left(\frac{\Delta-2}{2}\right) \ \left(\frac{s}{\Lambda^2}\right)^{2-\Delta/2}\, . \label{A4-glueballs-final}
\end{eqnarray}
The total cross section for this exclusive process at large $s$ and fixed scattering angle is
\begin{eqnarray}
\sigma_{\text{Total}} = \frac{1}{s} \ \text{Im}\mathcal{A}^{4d}_{4 \ {\text{glueballs}}}(p) \propto  \left(\frac{s}{\Lambda^2}\right)^{1-\Delta/2}\, . \label{Total-cross-section-glueballs}
\end{eqnarray}
In last part of this expression we have omitted the scattering angle dependence.

Now, let us consider the calculation of the angular integral of four scalar spherical harmonics on $S^5$ displayed in equation (\ref{Integral-4-scalars}). The spherical harmonics on $S^5$ are labelled by the set of positive integers $(l_1,l_2,l_3,l_4,l_5)$, satisfying $l_5 \ge l_4 \ge l_3 \ge l_2 \ge l_1 \ge 0$. We set $l_5 \equiv k$. The relation between the conformal dimension $\Delta$ and $k$ for scalars is $\Delta=k+4$ with $k \ge 0$. The explicit construction of the scalar spherical harmonics is presented in Appendix.

We may study some examples in order to appreciate how it works. Firstly, let us consider the irreducible representation of $SU(4)$ with $k=0$. This leads to only one scalar spherical harmonic\footnote{We switch to a more precise notation where we  explicitly write the integers $(l_1,l_2,l_3,l_4,l_5)$. Also, we use the label $(s)$ for scalar spherical harmonics.}
\begin{equation}
Y^{(s)}_{(0,0,0,0,0)}(\Omega)= \pi ^{-3/2} \,  , \label{Y00000}
\end{equation}
and the result of the integral (\ref{Integral-4-scalars}) with four $Y^{(s)}_{(0,0,0,0,0)}(\Omega)$'s is $\pi^{-3}$ as shown in table \ref{Table-1}. In this case the high-energy limit of the amplitude describes the scattering of four scalars in the $[0,0,0]$ irreducible representation of the $SU(4)_R$ $R$-symmetry group of $SU(N_c)$ ${\cal {N}}=4$ SYM theory. This corresponds to the case where all $\Delta_i=4$, i.e. $I_{4, 4, 4, 4}$, and $\Delta = 16$.

Next, let us consider the high-energy limit of two-to-two scattering amplitude for scalars corresponding to the $[0,1,0]$ irreducible representation of $SU(4)_R$, being $\Delta_i=5$. In this case there are 6 different scalar spherical harmonics on $S^5$, namely: $Y^{(s)}_{(1,0,0,0,0)}(\Omega)$, $Y^{(s)}_{(1,1,0,0,0)}(\Omega)$, $Y^{(s)}_{(1,1,1,0,0)}(\Omega)$, $Y^{(s)}_{(1,1,1,1,0)}(\Omega)$, $Y^{(s)}_{(1,1,1,1,1)}(\Omega)$ and $Y^{(s)}_{(1,1,1,1,-1)}(\Omega)$. There are many possible  combinations of the scalar spherical harmonics in this representation. There are 6 cases where the four scalars have the same set of $l_i$'s, namely:
\begin{eqnarray}
I_1=\int d\Omega_5 \ Y^{(s)*}_{(1,0,0,0,0)}(\Omega) \ Y^{(s)*}_{(1,0,0,0,0)}(\Omega) \ Y^{(s)}_{(1,0,0,0,0)}(\Omega) \ Y^{(s)}_{(1,0,0,0,0)}(\Omega) = \frac{9}{4 \pi^3} \ , \label{I1} \\
I_2=\int d\Omega_5 \ Y^{(s)*}_{(1,1,0,0,0)}(\Omega) \ Y^{(s)*}_{(1,1,0,0,0)}(\Omega) \ Y^{(s)}_{(1,1,0,0,0)}(\Omega) \ Y^{(s)}_{(1,1,0,0,0)}(\Omega) = \frac{9}{4 \pi^3} \ , \\
I_3=\int d\Omega_5 \ Y^{(s)*}_{(1,1,1,0,0)}(\Omega) \ Y^{(s)*}_{(1,1,1,0,0)}(\Omega) \ Y^{(s)}_{(1,1,1,0,0)}(\Omega) \ Y^{(s)}_{(1,1,1,0,0)}(\Omega) = \frac{9}{4 \pi^3} \ , \\
I_4=\int d\Omega_5 \ Y^{(s)*}_{(1,1,1,1,0)}(\Omega) \ Y^{(s)*}_{(1,1,1,1,0)}(\Omega) \ Y^{(s)}_{(1,1,1,1,0)}(\Omega) \ Y^{(s)}_{(1,1,1,1,0)}(\Omega) = \frac{9}{4 \pi^3} \ , \\
I_5=\int d\Omega_5 \ Y^{(s)*}_{(1,1,1,1,1)}(\Omega) \ Y^{(s)*}_{(1,1,1,1,1)}(\Omega) \ Y^{(s)}_{(1,1,1,1,1)}(\Omega) \ Y^{(s)}_{(1,1,1,1,1)}(\Omega) = \frac{3}{2 \pi^3} \ , \\
I_6=\int d\Omega_5 \ Y^{(s)*}_{(1,1,1,1,-1)}(\Omega) \ Y^{(s)*}_{(1,1,1,1,-1)}(\Omega) \ Y^{(s)}_{(1,1,1,1,-1)}(\Omega) \ Y^{(s)}_{(1,1,1,1,-1)}(\Omega) = \frac{3}{2 \pi^3} \ . \label{I6}
\end{eqnarray}
Thus, the corresponding scattering amplitudes are given by the generic schematic expression
\begin{equation}
\mathcal{A}^{4d}_{4 \ {\text{glueballs}}}(p) \propto  I_i \ \left(\frac{\Lambda}{p}\right)^{16} \, ,
\end{equation}
with the results of the integrals $I_i$'s with $i=1, \dots, 6$ given in equations (\ref{I1}) to (\ref{I6}), respectively. The exponent 16 comes from considering $\Delta=4 \times 5$ and write $s$ as $p^2$.

There are also many possibilities of two-to-two scattering of states belonging to the same or different scalar irreducible representations of $SU(4)_R$. In fact, the generic integral (\ref{Integral-4-scalars}) leads to selection rules for allowed states involved in these scattering processes. Some examples of the results of these integrals are presented in table \ref{Table-1}. In addition, there further restrictions that we will briefly comment in section 4.
\begin{table}[H]
\def\arraystretch{1.5}
\begin{center}
\begin{tabular}{|c|c|c|c|c|} 
 \hline
 & $Y_{(0,0,0,0,0)}^{(s)} Y_{(0,0,0,0,0)}^{(s)}$  & $Y_{(1,0,0,0,0)}^{(s)} Y_{(1,0,0,0,0)}^{(s)}$  & $Y_{(1,1,0,0,0)}^{(s)} Y_{(1,1,0,0,0)}^{(s)}$\\
\hline
$Y_{(0,0,0,0,0)}^{(s)*} Y_{(0,0,0,0,0)}^{(s)*}$ & $1/\pi^3$ &  $1/\pi^3$ &  $1/\pi^3$  \\
\hline
$Y_{(1,0,0,0,0)}^{(s)*} Y_{(1,0,0,0,0)}^{(s)*}$ & $1/\pi^3$ &     $9/(4\pi^3)$ & $3/(4\pi^3)$ \\
\hline
$Y_{(1,1,0,0,0)}^{(s)*} Y_{(1,1,0,0,0)}^{(s)*}$ & $1/\pi^3$ &     $3/(4\pi^3)$ & $9/(4\pi^3)$ \\
\hline
$Y_{(1,1,1,0,0)}^{(s)*} Y_{(1,1,1,0,0)}^{(s)*}$ & $1/\pi^3$ &     $3/(4\pi^3)$ & $3/(4\pi^3)$ \\
\hline
$Y_{(1,1,1,1,0)}^{(s)*} Y_{(1,1,1,1,0)}^{(s)*}$ & $1/\pi^3$ & $3/(4\pi^3)$ & $3/(4\pi^3)$ \\
\hline
$Y_{(1,1,1,1,1)}^{(s)*} Y_{(1,1,1,1,1)}^{(s)*}$ &    $0$    &    $0$          &  $0$         \\
\hline
$Y_{(1,1,1,1,-1)}^{(s)*} Y_{(1,1,1,1,-1)}^{(s)*}$ &   $0$   &   $0$          &  $0$         \\
\hline
\end{tabular}
\caption{\small  A few illustrative examples of the integral (\ref{Integral-4-scalars}) of four scalar spherical harmonics on $S^5$, leading to examples of selection rules.}
\label{Table-1}
\end{center}
\end{table}

Now, we consider the Regge limit $\tilde{s}>>|\tilde{t}|$ and $\tilde{s}\sim -\tilde{u}$ of the two-to-two glueballs scattering amplitude. The starting point is equation (\ref{K2-15}) setting $\theta \rightarrow 0$. Thus, the kinematic factor becomes
\begin{eqnarray}
K_{\text{closed}}^{\text{Regge}}(k_1,k_2,k_3,k_4) &=& \frac{9}{16} \tilde{s}^4  \ \Phi _1 \Phi_2 \Phi_3 \Phi_4  \, .
\end{eqnarray}
Next, we may use the Stirling's formula and obtain
\begin{eqnarray}
\prod_{\chi=\tilde{s},\tilde{t},\tilde{u}}\frac{\Gamma(-\alpha'\chi/4)}{\Gamma(1+\alpha'\chi/4)}
& \approx &-\left(\frac{\alpha'\tilde{s}}{4}\right)^{-2+\alpha'\tilde{t}/2}e^{2-\alpha'\tilde{t}/2}\frac{\Gamma(-\alpha'\tilde{t}/4)}{\Gamma(1+\alpha'\tilde{t}/4)} \, , 
\end{eqnarray}
having used the Euler's reflexion formula $\Gamma(z) \Gamma(1-z)=\pi/\sin(\pi z)$, which holds for any non-integer complex number $z$. Then, the Regge limit of the type II superstring theory scattering amplitude of four dilatons becomes
\begin{eqnarray}
  \mathcal{A}^{\text{Regge}}_{\text{string}} (\tilde{p})\approx\frac{9 \pi g_s^2 \alpha^{\prime 3}}{4}(\alpha'\tilde{s})^{2+\frac{1}{2}\alpha'\tilde{t}} \ (4 e)^{2-\alpha'\tilde{t}/2} \ \frac{\Gamma(-\alpha'\tilde{t}/4)}{\Gamma(1+\alpha'\tilde{t}/4)}  \ \Phi_1 \Phi_2 \Phi_3 \Phi_4 \, .
\end{eqnarray}
At this point equation (\ref{A4}) leads to the two-to-two glueballs scattering amplitude,
\begin{eqnarray}
\mathcal{A}^{4d \ \text{Regge}}_{4 \ {\text{glueballs}}}(p)&=&  9 \pi \ \frac{g_s^2 \ \alpha'^3}{R^6 r_0^{4-\Delta}} \ I_{\Delta_1,\Delta_2,\Delta_3,\Delta_4}\ \left(\prod_{i=1}^4 (\Delta_i-1)^{1/2}\right) \times \nonumber \\
&& \int_{r_0}^\infty dr \ r^{3-\Delta} \ (\alpha' \tilde{s})^4  \ \left(\frac{\alpha'\tilde{s}}{4}\right)^{-2+\alpha'\tilde{t}/2}e^{2-\alpha'\tilde{t}/2}\frac{\Gamma(-\alpha'\tilde{t}/4)}{\Gamma(1+\alpha'\tilde{t}/4)} \, .  \label{A4-glueball-Regge}
\end{eqnarray}
We can define $\tilde{t}=-|\tilde{t}|$ since in the Regge limit the Mandelstam variable $\tilde{t} \leq 0$. Moreover, for $|\tilde{t}| \ll 1$ the above quotient of Gamma functions leads to a factor $1/|\tilde{t}|$. Therefore, we must solve the radial integral
\begin{eqnarray}
I_r & \approx & \int_{r_0}^\infty dr \ r^{5-\Delta} \ \left(\frac{\alpha' R^2 s}{r^2}\right)^{2+\frac{1}{2} \alpha' R^2 t/r^2} \ (4 \ e)^{2-\frac{1}{2} \alpha' R^2 t/r^2} \left(\frac{4}{\alpha' R^2 |t|}\right) \, ,
\end{eqnarray}
which can be solved by using the saddle-point approximation.
Thus, we rewrite this radial integral as
\begin{eqnarray}
I_r & \approx & \int_{r_0}^\infty dr \ \exp[h(r)] \, ,
\end{eqnarray}
where
\begin{eqnarray}
h(r) & = & \log\left(r^{5-\Delta} \ \left(\frac{\alpha' R^2 s}{r^2}\right)^{2+\frac{1}{2} \alpha' R^2 t/r^2} \ (4 \ e)^{2-\frac{1}{2} \alpha' R^2 t/r^2} \left(\frac{4}{\alpha' R^2 |t|}\right)\right) \, .
\end{eqnarray}
Then, the condition $dh(r^*)/r=0$ is satisfied by the saddle point $r^*$
\begin{eqnarray}
r^*=R\sqrt{\alpha'|t|} \sqrt{\frac{\log\left(\frac{s}{|t|}\right)+\log\left(\frac{\Delta-1}{4}\right)}{\Delta-1}} \, .
\end{eqnarray}
The dominant value of the radial coordinate comes from 
\begin{eqnarray} 
r'&=& {\text{min}} \left(\Lambda R^2, r^* \right)
\end{eqnarray}
leading to the Regge behavior if the condition
\begin{eqnarray}
\sqrt{\hat{\alpha}'|t|} \sqrt{\frac{\log\left(\frac{s}{|t|}\right)+\log\left(\frac{\Delta-1}{4}\right)}{\Delta-1}} > 1 \, 
\end{eqnarray}
is satisfied. Then, equation (\ref{A4-glueball-Regge}) becomes
\begin{eqnarray}
\mathcal{A}^{4d \ \text{Regge}}_{4 \ {\text{glueballs}}}(p)& \approx&  \frac{36 \sqrt{4 \pi g_s N_c}}{\pi^{1/2} N_c^2} \ I_{\Delta_1,\Delta_2,\Delta_3,\Delta_4}\ \left(\prod_{i=1}^4 (\Delta_i-1)^{1/2}\right) \times \nonumber \\
&& \frac{2^{\frac{1}{2}+\hat{\alpha'} |t|} e^{2+\frac{\hat{\alpha'}}{2} |t|} (\hat{\alpha'} s)^{2+\frac{\hat{\alpha'}}{2} t}}{\hat{\alpha'} |t| \sqrt{|\Delta-1 +\hat{\alpha'} |t| (3 \log 4-2) - 3 \hat{\alpha'} |t| \log(\hat{\alpha'} s) |}} \, .  \label{A4-glueball-Regge-2}
\end{eqnarray}
If we consider that $\hat{\alpha'} |t| \approx 1$ we obtain
\begin{eqnarray}
\mathcal{A}^{4d \ \text{Regge}}_{4 \ {\text{glueballs}}}(p)&\approx&  \frac{36 \cdot 2^{3/2} \sqrt{4 \pi g_s N_c} \ e^{5/2}}{\pi^{1/2} N_c^2} \ I_{\Delta_1,\Delta_2,\Delta_3,\Delta_4}\ \left(\prod_{i=1}^4 (\Delta_i-1)^{1/2}\right) \times \nonumber \\
&& \frac{ (\hat{\alpha'} s)^{2+\frac{\hat{\alpha'}}{2} t}}{\sqrt{|\Delta-1 + (3 \log 4-2) - 3  \log(\hat{\alpha'} s) |}} \, .  \label{A4-glueball-Regge-2}
\end{eqnarray}
The factor $(\hat{\alpha'}s)^{2+\frac{\hat{\alpha'} t}{2}}$ can be interpreted as a Regge trajectory $j(t)=2+\frac{\hat{\alpha'}t}{2}=2+\frac{t}{2\sqrt{\lambda}\Lambda^2}$, obtained from the Reggeization of a graviton propagating between the two glueballs in the $t$-channel.

%
%
\section{High-energy limit of four spin-1/2 fermions scattering amplitude}
%
%

In this section we focus on the calculation of the high-energy behavior of the four-dilatino scattering amplitude which we obtained from type IIB superstring theory in our work \cite{MPS}. In particular, we carry out the calculation of the four-dimensional scattering amplitude of four spin-1/2 fermions in the dual SYM theory, considering the fixed scattering angle and the Regge limit situations. The starting point in \cite{MPS} is the Kawai-Lewellen-Tye (KLT) relation \cite{Kawai:1985xq} which indicates that tree-level closed string scattering amplitudes can be written as the product of two open string scattering amplitudes. For the kinematic factor 
we have:
\begin{equation}
K_{\text{closed}}^{\text{4-dilatino}}(\tilde{1},\tilde{2}, \tilde{3},\tilde{4})= K_{\text{open}}^{\text{fermionic}}(\tilde{1},\tilde{2}, \tilde{3},\tilde{4})\otimes \frac{1}{16} K_{\text{open}}^{\text{bosonic}}(1,2,4,3) \, .\label{Kclosed4dilatinos} 
\end{equation}
In this equation we define $K_{\text{open}}^{\text{bosonic}}(1,2,4,3) \equiv K_{\text{open}}(k_1,k_2,k_4,k_3)$ which is given in equation (\ref{Kopen4}). Here we have simplified the notation replacing $k_j \rightarrow j$. Notice that we only consider massless string states, thus $\tilde{s}+\tilde{t}+\tilde{u}=0$ holds. As specified by the KLT relations in $K_{\text{open}}^{\text{bosonic}}(1,2,4,3)$ we exchange strings 3 and 4.

The open string kinematic factor for four spin-1/2 fermions is 
\cite{Green:1987sp}
\begin{eqnarray}
K_{\text{open}}^{\text{fermionic}}(\tilde{1},\tilde{2},\tilde{3},\tilde{4} )&=& -\frac{1}{8} \tilde{s} \bar{u}_2 \Gamma^{M'} u_{3} \bar{u}_1\Gamma_{M'} u_4 +  \frac{1}{8} \tilde{t} \bar{u}_1 \Gamma^{M'} u_{2} \bar{u}_4  \Gamma_{M'} u_3 \, . \label{Kopen4fermions}
\end{eqnarray} 
Spinor polarizations are indicated as $u_i$'s. $\Gamma_M$ represents the Dirac gamma matrices in ten-dimensional Mikowski spacetime. We may define the dilatino as
\begin{eqnarray}
u_i^\alpha\otimes \zeta_i^M =(\Gamma^M)^\alpha_\beta \ \lambda_i^\beta    \, ,
\end{eqnarray}
being $\lambda$ the ten-dimensional 32 component dilatino in type IIB superstring theory, and $\text{M,N}, \dots$ are ten-dimensional Lorentz indices, while $\alpha, \beta, \dots$ are spinor indices. The resulting amplitude $\mathcal{A}_{\text{string}}^{10d,{\text{4 dilatinos}}}(\tilde{p})$ obtained in \cite{MPS} contains 30 terms, being the structure of each one very complicated (mainly due to the presence of 6 Dirac gamma matrices of 32$\times$32 components). For simplicity we opt to address a couple of representative terms in order to calculate the high-energy behavior of the four-dimensional scattering amplitude of four spin-1/2 fermions in the dual gauge theory. These terms will display the leading and subleading behavior in powers of $s$. Therefore, let us consider the following expression
\begin{eqnarray}
K_{\text{closed}}^{\text{4-dilatino}}(\tilde{1},\tilde{2}, \tilde{3},\tilde{4})
&=& \frac{\tilde{s}^2\tilde{t}}{512}\bar{\lambda}_2\Gamma^N\Gamma^P\Gamma^M\lambda_3\bar{\lambda}_1\Gamma_M\Gamma_P\Gamma_N\lambda_4 \nonumber \\
& &-\frac{\tilde{s}\tilde{t}^2}{512}\bar{\lambda}_1\Gamma^N\Gamma^P\Gamma^M\lambda_2\bar{\lambda}_4\Gamma_M\Gamma_P\Gamma_N\lambda_3+ \dots \, ,
\end{eqnarray}
where $\dots$ indicates the rest of terms obtained in \cite{MPS} that we omit in the preceding equation. Taking into account that
$(\Gamma^0)^2=-1$, $(\Gamma^i)^2=1$ and $\Gamma^M\Gamma_M=10$, we can do some algebraic work on the kinematic factor obtaining
\begin{eqnarray}
K_{\text{closed}}^{\text{4-dilatino}}(\tilde{1},\tilde{2}, \tilde{3},\tilde{4})&=&\frac{\tilde{s}^2\tilde{t}}{512}[15\bar{\lambda}_2\Gamma^M\lambda_3\bar{\lambda}_1\Gamma_M\lambda_4+\bar{\lambda}_2\Gamma^{N(\neq M \neq P)}\Gamma^{P(\neq M \neq N)}\Gamma^{M(\neq N \neq P)}\lambda_3\nonumber\\
& \times & \bar{\lambda}_1\Gamma_{M(\neq N \neq P)}\Gamma_{P(\neq M \neq N)}\Gamma_{N(\neq M \neq P)}\lambda_4] \nonumber \\
&-&\frac{\tilde{s}\tilde{t}^2}{512}[15\bar{\lambda}_1\Gamma^M\lambda_2\bar{\lambda}_4\Gamma_M\lambda_3+\bar{\lambda}_1\Gamma^{N(\neq M \neq P)}\Gamma^{P(\neq M \neq N)}\Gamma^{M(\neq N \neq P)}\lambda_2\nonumber\\
& \times & \bar{\lambda}_4\Gamma_{M(\neq N \neq P)}\Gamma_{P(\neq M \neq N)}\Gamma_{N(\neq M \neq P)}\lambda_3] +  \dots \, ,
\end{eqnarray}
where $N(\neq M \neq P)$ simply means that the ten-dimensional Lorentz index $N$ cannot be equal to $M$ nor $P$.

Since we are interested in the calculation of the scattering amplitude in the dual gauge theory, we have to carry out the integration on the radial and angular coordinates on AdS$_5 \times S^5$. For convenience we express the AdS$_5$ metric in the following way
\begin{eqnarray}
ds^2=R^2\frac{\eta_{\mu\nu} dx^{\mu} dx^{\nu} + dz^2}{z^2} = g_{\mu\nu}dx^{\mu} dx^{\nu} + g_{zz} dz^2 = g_{ab} dx^a dx^b \, ,
\end{eqnarray}
where $a, b, c=0, \dots, 4$ are in the AdS$_5$ space. On the other hand, the range of the $z$ coordinate is $0 \leq z \leq z_0$, being $z_0$ the IR cut-off. The relations between flat and curved space coordinates in five dimensions are given by
\begin{eqnarray}
\eta_{\hat{a}\hat{b}}e^{\hat{a}}_{a}e^{\hat{b}}_b=g_{ab} \, ,
\,\,\,\,\,\,\,\,\, {\text{with}} \,\,\,\,\,\,\,\,\, 
e^{\hat{a}}_{a}=\sqrt{g_{aa}}\delta_a^{\hat{a}} \, , \,\,\,\,\,\,\,\,\, {\text{and}} \,\,\,\,\,\,\,\,\, 
e^{a}_{\hat{a}}=\frac{1}{\sqrt{g_{\hat{a}\hat{a}}}}\delta_{\hat{a}}^a \, ,
\end{eqnarray}
where $\hat{a},\hat{b},\hat{c}=0,1,2,3,4$ are the five-dimensional flat spacetime Lorentz indices. In order to further simplify the analysis we focus on the AdS$_5$ part. Thus, let us study terms of the following form, which have coordinates of type $\hat{a},\hat{b},\hat{c}$,
\begin{eqnarray}
 K_{\text{closed}}^{\text{4-dilatino}}(\tilde{1},\tilde{2}, \tilde{3},\tilde{4})   & = &\frac{\tilde{s}^2\tilde{t}}{512}[15\bar{\lambda}_2\Gamma^{\hat{a}}\lambda_3\bar{\lambda}_1\Gamma_{\hat{a}}\lambda_4+\bar{\lambda}_2\Gamma^{\hat{b}(\neq \hat{a} \neq \hat{c})}\Gamma^{\hat{c}(\neq \hat{a} \neq \hat{b})}\Gamma^{\hat{a}(\neq \hat{b} \neq \hat{c})}\lambda_3\nonumber\\
&\times & \bar{\lambda}_1\Gamma_{\hat{a}(\neq \hat{b} \neq \hat{c})}\Gamma_{\hat{c}(\neq \hat{a} \neq \hat{b})}\Gamma_{\hat{b}(\neq \hat{a} \neq \hat{c})}\lambda_4] \nonumber \\
&-&\frac{\tilde{s}\tilde{t}^2}{512}[15\bar{\lambda}_1\Gamma^{\hat{a}}\lambda_2\bar{\lambda}_4\Gamma_{\hat{a}}\lambda_3+\bar{\lambda}_1\Gamma^{\hat{b}(\neq \hat{a} \neq \hat{c})}\Gamma^{\hat{c}(\neq \hat{a} \neq \hat{b})}\Gamma^{\hat{a}(\neq \hat{b} \neq \hat{c})}\lambda_2\nonumber\\
&\times & \bar{\lambda}_4\Gamma_{\hat{a}(\neq \hat{b} \neq \hat{c})}\Gamma_{\hat{c}(\neq \hat{a} \neq \hat{b})}\Gamma_{\hat{b}(\neq \hat{a} \neq \hat{c})}\lambda_3]+ \dots \, .
\end{eqnarray}

Next step is to consider equation (\ref{A4}) for the case  of four dilatinos, leading to the scattering amplitude $\mathcal{A}_{\text{4 spin-1/2 fermions}}^{4d}(p)$, which implies to carry out the radial integral (after replacing $r=R^2/z$) and the angular one, considering the AdS$_5 \times S^5$ background. For that purpose, firstly we must consider the dilatinos wave-functions $\lambda_j(y, \Omega)$. Thus, we  have to plug the following ansatz in equation (\ref{A4})
\begin{eqnarray}
\widetilde{K}_{\text{closed}}^{\text{4-dilatino}}(\tilde{1},\tilde{2}, \tilde{3},\tilde{4})   & = &
\frac{\tilde{s}^2\tilde{t}}{512}[15\bar{\lambda}_2(y,\Omega)\tilde{\Gamma}^a\lambda_3(y,\Omega)\bar{\lambda}_1(y,\Omega)\tilde{\Gamma}_a\lambda_4(y,\Omega) \nonumber \\
&+&\bar{\lambda}_2(y,\Omega)\tilde{\Gamma}^{b(\neq a \neq c)}\tilde{\Gamma}^{c(\neq a \neq b)}\tilde{\Gamma}^{a(\neq b \neq c)}\lambda_3(y,\Omega)\nonumber\\
&\times&\bar{\lambda}_1(y,\Omega)\tilde{\Gamma}_{a(\neq b \neq c)}\tilde{\Gamma}_{c(\neq a \neq b)}\tilde{\Gamma}_{b(\neq a \neq c)}\lambda_4(y,\Omega)]\nonumber\\
&-&\frac{\tilde{s}\tilde{t}^2}{512}[15\bar{\lambda}_1(y,\Omega)\tilde{\Gamma}^a\lambda_2(y,\Omega)\bar{\lambda}_4(y,\Omega)\tilde{\Gamma}_a\lambda_3(y,\Omega)\nonumber\\
&+&\bar{\lambda}_1(y,\Omega)\tilde{\Gamma}^{b(\neq a \neq c)}\tilde{\Gamma}^{c(\neq a \neq b)}\tilde{\Gamma}^{a(\neq b \neq c)}\lambda_2(y,\Omega)\bar{\lambda}_4(y,\Omega) \nonumber \\
&\times& \tilde{\Gamma}_{a(\neq b \neq c)}\tilde{\Gamma}_{c(\neq a \neq b)}\tilde{\Gamma}_{b(\neq a \neq c)}\lambda_3(y,\Omega)]+ \dots \, ,
\end{eqnarray}
where $\widetilde{K}_{\text{closed}}^{\text{4-dilatino}}(\tilde{1},\tilde{2}, \tilde{3},\tilde{4})$ indicates that this is an ansatz using a local approximation analogous to the one proposed by Polchinski and Strassler for the dilaton scattering amplitude in curved space. Note that $y=(r, x^{\mu})$ and $\Omega$ are the $S^5$ coordinates. In addition, the curved space-time $\tilde{\Gamma}_a$'s are in the following representation 
\begin{eqnarray}
\tilde{\Gamma}^a=\sigma^1\otimes I_4 \otimes \tilde{\gamma}^a \, ,
\end{eqnarray}
where $\tilde{\gamma}^a$ are the Dirac matrices in AdS$_5$.

On the other hand, in flat space-time the Dirac gamma matrices satisfy
\begin{eqnarray}
\gamma^{\mu}\gamma^{\nu}\gamma^{\rho}=\eta^{\mu\nu}\gamma^{\rho}+\eta^{\nu\rho}\gamma^{\mu}-\eta^{\mu\rho}\gamma^{\nu}-i\epsilon^{\sigma\mu\nu\rho}\gamma_{\sigma}\gamma^5 \, , \label{gamma-flat-properties}
\end{eqnarray}
with $\epsilon^{\sigma\mu\nu\rho}\epsilon_{\alpha\mu\nu\rho}=6 \ \delta^{\sigma}_{\alpha}$. They also satisfy the Clifford algebra $\lbrace \gamma^{\mu},\gamma^{\nu}\rbrace = 2\eta^{\mu\nu}$, while $\gamma^5=i\gamma^0\gamma^1\gamma^2\gamma^3$.

We consider the following representation
\begin{eqnarray}
\gamma^{\mu}=\begin{pmatrix} 
0 &-i\sigma^{\mu} \\ -i\bar{\sigma}^{\mu} & 0
    \end{pmatrix} \, ,
\end{eqnarray}
in terms of the Pauli matrices
\begin{eqnarray}
\sigma^1=\begin{pmatrix}0 & 1 \\ 1 & 0 
\end{pmatrix}, \,\,\,\,\,\,\,\,\,\,\,
\sigma^2=\begin{pmatrix}0 & -i \\ i & 0\end{pmatrix}, \,\,\,\,\,\,\,\,\,\,\,
\sigma^3=\begin{pmatrix}1 & 0\\ 0& -1
\end{pmatrix} \, .
\end{eqnarray}
The relations between flat and curved space five-dimensional gamma matrices are given by
\begin{eqnarray}
    \widetilde{\gamma}_a=e_a^{\hat{a}}\gamma_{\hat{a}}=\sqrt{g_{aa}}\gamma_{\hat{a}}=\frac{R}{z}\gamma_{\hat{a}} \, ,
\end{eqnarray}
and
\begin{eqnarray}
\widetilde{\gamma}^a=e_{\hat{a}}^{a}\gamma^{\hat{a}}=\frac{1}{\sqrt{g_{\hat{a}\hat{a}}}}\gamma^{\hat{a}}=\frac{z}{R}\gamma^{\hat{a}} \, .
\end{eqnarray}
The right-handed dilatino has only 16 non-vanishing components and can be written as follows   
\begin{eqnarray}
\lambda=\begin{pmatrix}
0 \\ \lambda' 
\end{pmatrix} \, ,  
\end{eqnarray}
with the following Kaluza-Klein decomposition in AdS$_5 \times S^5$   
\begin{eqnarray}
\lambda'_i(y,\Omega)=\lambda_i(y)\otimes Y_i(\Omega) \, , \label{dilatino-1}
\end{eqnarray}
where $\lambda_i(y)$ represents an AdS$_5$ spinor of four components while $Y_i(\Omega)$ is an spinor spherical harmonic on $S^5$. Thus, the kinematic factor for the four dilatinos in the curved-space ansatz becomes
\begin{eqnarray}
& & \widetilde{K}_{\text{closed}}^{\text{4-dilatino}}(\tilde{1},\tilde{2}, \tilde{3},\tilde{4}) = \nonumber \\
&& \frac{\tilde{s}^2\tilde{t}}{512}[15\overline{\lambda_2(y)\otimes Y_2(\Omega)}(I_4\otimes\tilde{\gamma}^a)\lambda_3(y)\otimes Y_3(\Omega)\overline{\lambda_1(y)\otimes Y_1(\Omega)}(I_4\otimes\tilde{\gamma}_a)\lambda_4(y)\otimes Y_4(\Omega)\nonumber\\
&+&\overline{\lambda_2(y)\otimes Y_2(\Omega)}(I_4\otimes\tilde{\gamma}^{b(\neq a \neq c)})(I_4\otimes\tilde{\gamma}^{c(\neq a \neq b)})(I_4\otimes\tilde{\gamma}^{a(\neq b \neq c)})\lambda_3(y)\otimes Y_3(\Omega)\nonumber\\
&\times & \overline{\lambda_1(y)\otimes Y_1(\Omega)}(I_4\otimes\tilde{\gamma}_{a(\neq b \neq c)})(I_4\otimes\tilde{\gamma}_{c(\neq a \neq b)})(I_4\otimes\tilde{\gamma}_{b(\neq a \neq c)})\lambda_4(y)\otimes Y_4(\Omega)]\nonumber\\
&-&\frac{\tilde{s}\tilde{t}^2}{512}[15\overline{\lambda_1(y)\otimes Y_1(\Omega)}(I_4\otimes\tilde{\gamma}^a)\lambda_2(y)\otimes Y_2(\Omega)\overline{\lambda_4(y)\otimes Y_4(\Omega)}(I_4\otimes\tilde{\gamma}_a)\lambda_3(y)\otimes Y_3(\Omega)\nonumber\\
&+&\overline{\lambda_1(y)\otimes Y_1(\Omega)}(I_4\otimes\tilde{\gamma}^{b(\neq a \neq c)})(I_4\otimes\tilde{\gamma}^{c(\neq a \neq b)})(I_4\otimes\tilde{\gamma}^{a(\neq b \neq c)})\lambda_2(y)\otimes Y_2(\Omega)\nonumber\\
&\times&\overline{\lambda_4(y)\otimes Y_4(\Omega)}(I_4\otimes\tilde{\gamma}_{a(\neq b \neq c)})(I_4\otimes\tilde{\gamma}_{c(\neq a \neq b)})(I_4\otimes\tilde{\gamma}_{b(\neq a \neq c)})\lambda_3(y)\otimes Y_3(\Omega)]+ \dots  \, . \nonumber \\
&&
\end{eqnarray}
An additional simplification can be considered if we factorize out the products of the four spinor spherical harmonics of each term in the previous expression. Since we are only interested in the high-energy behavior, i.e. the powers of $s$ in the four-dimensional scattering amplitude, this approximation is reasonable. Then, the kinematic factor reads
\begin{eqnarray}
&& \widetilde{K}_{\text{closed}}^{\text{4-dilatino}}(\tilde{1},\tilde{2}, \tilde{3},\tilde{4}) \approx \nonumber \\
&& \frac{\tilde{s}^2\tilde{t}}{512}[15\bar{\lambda}_2(y)\tilde{\gamma}^a \lambda_3(y)\bar{\lambda}_1(y) \tilde{\gamma}_a \lambda_4(y)\nonumber\\
&+&\bar{\lambda}_2(y)(\tilde{\gamma}^{b(\neq a \neq c)})(\tilde{\gamma}^{c(\neq a \neq b)})(\tilde{\gamma}^{a(\neq b \neq c)})\lambda_3(y)\bar{\lambda}_1(y)(\tilde{\gamma}_{a(\neq b \neq c)})(\tilde{\gamma}_{c(\neq a \neq b)})(\tilde{\gamma}_{b(\neq a \neq c)})\lambda_4(y)]\nonumber\\
&-&\frac{\tilde{s}\tilde{t}^2}{512}[15\bar{\lambda}_1(y)\tilde{\gamma}^a\lambda_2(y)\bar{\lambda}_4(y)\tilde{\gamma}_a\lambda_3(y)+\bar{\lambda}_1(y)(\tilde{\gamma}^{b(\neq a \neq c)})(\tilde{\gamma}^{c(\neq a \neq b)})(\tilde{\gamma}^{a(\neq b \neq c)})\lambda_2(y)\nonumber\\
&\times&\bar{\lambda}_4(y)(\tilde{\gamma}_{a(\neq b \neq c)})(\tilde{\gamma}_{c(\neq a \neq b)})(\tilde{\gamma}_{b(\neq a \neq c)})\lambda_3(y)]+ \dots \, .
\end{eqnarray}
It is very easy to show that flat-space and curved-space five-dimensional gamma matrices lead to the same kinematic factor for the four fermions kinematic factor. Then, using the properties (\ref{gamma-flat-properties}), the ten-dimensional four-dilatino scattering amplitude with the curved-space ansatz becomes
\begin{eqnarray}
& ~ & \widetilde{\mathcal{A}}_{\text{string (curved)}}^{10d,{\text{4 dilatinos}}}(\tilde{p}) \approx -\pi  \kappa_{10}^2  \alpha'^3 \
\widetilde{K}_{\text{closed}}^{\text{4-dilatino}}(\tilde{1},\tilde{2}, \tilde{3},\tilde{4})
)\prod_{\chi=\tilde{s},\tilde{t},\tilde{u}}\frac{\Gamma(-\alpha'\chi/4)}{\Gamma(1+\alpha'\chi/4)} \nonumber \\
&=& \frac{\pi g_s^2 \alpha'^4 (\alpha'\tilde{s})^3(1-cos\theta)}{256}\lbrace[15\bar{\lambda}_2(y)\gamma^{\mu}\lambda_3(y)\bar{\lambda}_1(y)\gamma_{\mu}\lambda_4(y) \nonumber\\
&+& 15\bar{\lambda}_2(y)\gamma^{5}\lambda_3(y)\bar{\lambda}_1(y)\gamma^{5}\lambda_4(y)+6\bar{\lambda}_2(y)\gamma_{\sigma}\gamma^5\lambda_3(y)\bar{\lambda}_1(y)\gamma^{\sigma}\gamma^5\lambda_4(y)\nonumber\\
&+&\bar{\lambda}_2(y)(\gamma^{\nu( \neq \rho)})(\gamma^{\rho(\neq z \neq \nu)})\gamma^{z}\lambda_3(y)\bar{\lambda}_1(y)\gamma^{5}(\gamma_{\rho( \neq \nu)})(\gamma_{\nu( \neq \rho)})\lambda_4(y)\nonumber\\
&+&\bar{\lambda}_2(y)\gamma^{5}(\gamma^{\rho(\neq \mu )})(\gamma^{\mu( \neq \rho)})\lambda_3(y)\bar{\lambda}_1(y)(\gamma_{\mu( \neq \rho)})(\gamma_{\rho(\neq \mu )})\gamma^{5}\lambda_4(y)\nonumber\\
&+&\bar{\lambda}_2(y)(\gamma^{\nu(\neq \mu )})\gamma^{5}(\gamma^{\mu(\neq \nu )})\lambda_3(y)\bar{\lambda}_1(y)(\gamma_{\mu(\neq \nu )})\gamma^{5}(\gamma_{\nu(\neq \mu )})\lambda_4(y)]\nonumber\\
&+&\frac{(1-cos\theta)}{2}[15\bar{\lambda}_1(y)\gamma^{\mu}\lambda_2(y)\bar{\lambda}_4(y)\gamma_{\mu}\lambda_3(y)+15\bar{\lambda}_1(y)\gamma^{5}\lambda_2(y)\bar{\lambda}_4(y)\gamma^{5}\lambda_3(y)\nonumber\\
&+&6\bar{\lambda}_1(y)\gamma_{\sigma}\gamma^5\lambda_2(y)\bar{\lambda}_4(y)\gamma^{\sigma}\gamma^5\lambda_3(y)\nonumber\\
&+&\bar{\lambda}_1(y)(\gamma^{\nu( \neq \rho)})(\gamma^{\rho( \neq \nu)})\gamma^{5}\lambda_2(y)\bar{\lambda}_4(y)\gamma^{5}(\gamma_{\rho( \neq \nu)})(\gamma_{\nu( \neq \rho)})\lambda_3(y)\nonumber\\
&+&\bar{\lambda}_1(y)\gamma^{5}(\gamma^{\rho(\neq \mu)})(\gamma^{\mu( \neq \rho)})\lambda_2(y)\bar{\lambda}_4(y)(\gamma_{\mu( \neq \rho)})(\gamma_{\rho(\neq \mu)})\gamma^{5}\lambda_3(y)\nonumber\\
&+&\bar{\lambda}_1(y)(\gamma^{\nu(\neq \mu )})\gamma^{5}(\gamma^{\mu(\neq \nu )})\lambda_2(y)\bar{\lambda}_4(y)(\gamma_{\mu(\neq \nu )})\gamma^{5}(\gamma_{\nu(\neq \mu )})\lambda_3(y)]+ \dots \rbrace \nonumber \\
&\times& \prod_{\chi=\tilde{s},\tilde{t},\tilde{u}}\frac{\Gamma(-\alpha'\chi/4)}{\Gamma(1+\alpha'\chi/4)}\, . 
\end{eqnarray}
Notice that $\gamma^z \equiv \gamma^5$. Then, from equation (\ref{A4}) in the present case we have
\begin{eqnarray}
\mathcal{A}^{4 d}_{\text{4 fermions}}(p)=\int_{r_0}^{\infty}dr \ \int_{S^5} d\Omega_5 \ \sqrt{-g} \ \widetilde{\mathcal{A}}_{\text{string (curved)}}^{10d,{\text{4 dilatinos}}}(\tilde{p}) \, . \label{A4-dil}
\end{eqnarray}
As in the case of glueball scattering, since we are interested in the high-energy behavior of the dual gauge theory, it is enough to study the asymptotic behavior of the dilatino wave-function. The general solution for the 16-component dilatino is
\begin{eqnarray}
 \lambda'_i(y,\Omega) =c_ie^{i P \cdot x}f'_i(r/r_0)Y_i(\Omega) \equiv c_i e^{i P \cdot x} \Psi_i(r,\Omega) \, , \label{dilatino-0}
\end{eqnarray}
where the four-dimensional dilatino $\lambda(y)$ from (\ref{dilatino-1}) satisfies the following AdS$_5$ Dirac's equation
\begin{eqnarray}
\left(z\gamma^m\partial_m-2\gamma^5+m_k\right)\lambda_i(y)=0 \, ,
\end{eqnarray}
where $m_k=k+\frac{3}{2}$, which has the following general solution
\begin{eqnarray}
\lambda_i(y)= Ce^{iP \cdot x}z^{5/2}\left[ a^+(P)J_{mR-1/2}(Mz) + a^-(P)J_{mR+1/2}(Mz)\right] \, , \label{5-d-spinor-solution}
\end{eqnarray}
with the Dirac's equation in four dimensions $i\gamma^{\mu}P_{\mu}u_{\sigma}=Mu_{\sigma}$. In this way, from the five-dimensional Dirac spinors $a_{\pm}$ we can construct four-dimensional Dirac spinors $u_{\sigma}$ satisfying the Einstein dispersion relation $P^2=-M^2$. Also there are the usual four-dimensional projectors $P_{\pm}=\frac{(1 \pm \gamma^5)}{2}$. Acting with $P_+$ on both sides of $i\gamma^{\mu}P_{\mu}u_{\sigma}=Mu_{\sigma}$ and doing a trivial algebra we obtain
\begin{eqnarray}
a_+&  = & \frac{i\gamma^{\mu}P_{\mu}}{M}a_- \, .
\end{eqnarray}
The solution (\ref{5-d-spinor-solution}) can be rewritten as
\begin{eqnarray}
\lambda_i(y)= Ce^{i P \cdot x} z^{5/2}\left[ P_+J_{mR-1/2}(Mz) + P_-J_{mR+1/2}(Mz)\right]u_\sigma \, . \label{dilatino-sol-2}
\end{eqnarray}
Since we are interested in the high-energy behavior of the scattering amplitude, we have to consider large $r$ values (recall that $r=R^2/z$), thus $p \sim\sqrt{s} \sim\frac{r}{R^2}$ must be large. Therefore, for the Bessel functions we have $J(Mz)=J(M\frac{R^2}{r})\sim J(\frac{M}{\sqrt{s}})$. In fact, for small values of the argument 
\begin{eqnarray}
J_{m R-1/2}(M z)=\frac{(M z)^{m R-1/2}}{2^{m R-1/2}(m R-1/2)!} \, ,
\end{eqnarray}
and similarly for $J_{m+1/2}(Mz)$. Thus, from equation (\ref{dilatino-sol-2})  expressed in terms of the radial coordinate $r$ we have,
\begin{eqnarray}
\lambda_i^{\sigma}(y)&\approx& e^{i P \cdot x}\frac{c_i}{R^{9/2}\Lambda^{3/2}(\Delta_i-5/2)!}(r/r_0)^{-\Delta_i}\left[ P_{+} + \frac{M_iR^2}{2r(\Delta_i-3/2)}P_-\right]u_{i}^{\sigma} \, , 
\end{eqnarray}
and
\begin{eqnarray}
    \bar{\lambda}_i^{\sigma}(y)&\approx& e^{-i P \cdot x}\frac{c_i^*}{R^{9/2}\Lambda^{3/2}(\Delta_i-5/2)!}(r/r_0)^{-\Delta_i} \ \bar{u}_{i}^{\sigma}\left[ P_{-} + \frac{M_iR^2}{2r(\Delta_i-3/2)}P_+\right] \, .
\end{eqnarray}
Recall that the spinors $\lambda_i^{\sigma}(y)$ come from the Kaluza-Klein decomposition of the 32-component Majorana-Weyl right-handed spinors after spontaneous compactification of type IIB supergravity on AdS$_5 \times S^5$. They are the dual fields to the $\mathcal{O}^{(6)}_k(x)$ operators of $\mathcal{N}=4$ SYM theory. The relation between the corresponding twist and the conformal dimension is $\tau=\Delta-1/2=m R+3/2$, with $\Delta=k+7/2$. The constant $c_i$ is obtained from the normalization condition on (\ref{dilatino-0}) leading to
\begin{eqnarray}
c_i=\sqrt{2 \Lambda R(\Delta_i-1)} \ (\Delta_i-5/2)! \, .
\end{eqnarray}
The four-dimensional scattering amplitude of four spin-1/2 fermions (\ref{A4-dil}) reads
\begin{eqnarray}
&& \mathcal{A}^{4 d}_{\text{4 fermions}}(p) \approx
-\left(\frac{128e^2}{\alpha'^3}\right)\frac{\pi g_s^2 R^2 \alpha'^4(1-\cos\theta)}{256} \nonumber \\
&\times& \int_{r_0}^{\infty}dr \ r^3 \ (-1)^{\alpha' \tilde{s}/2} (\alpha'\tilde{s})^3\lbrace[15\bar{\lambda}_2(y)\gamma^{\mu}\lambda_3(y)\bar{\lambda}_1(y)\gamma_{\mu}\lambda_4(y)\nonumber\\
&+&15\bar{\lambda}_2(y)\gamma^{5}\lambda_3(y)\bar{\lambda}_1(y)\gamma^{5}\lambda_4(y)+6\bar{\lambda}_2(y)\gamma_{\sigma}\gamma^5\lambda_3(y)\bar{\lambda}_1(y)\gamma^{\sigma}\gamma^5\lambda_4(y)\nonumber\\
&+&\bar{\lambda}_2(y)\gamma^{\nu( \neq \rho)}\gamma^{\rho(\neq z \neq \nu)}\gamma^{z}\lambda_3(y)\bar{\lambda}_1(y)\gamma^{5}\gamma_{\rho( \neq \nu)}\gamma_{\nu( \neq \rho)}\lambda_4(y)\nonumber\\
&+&\bar{\lambda}_2(y)\gamma^{5}\gamma^{\rho(\neq \mu )}\gamma^{\mu( \neq \rho)}\lambda_3(y)\bar{\lambda}_1(y)\gamma_{\mu( \neq \rho)}\gamma_{\rho(\neq \mu )}\gamma^{5})\lambda_4(y)\nonumber\\
&+&\bar{\lambda}_2(y)\gamma^{\nu(\neq \mu )}\gamma^{5}\gamma^{\mu(\neq \nu )}\lambda_3(y)\bar{\lambda}_1(y)\gamma_{\mu(\neq \nu )}\gamma^{5}\gamma_{\nu(\neq \mu )}\lambda_4(y)]\nonumber\\
&+&\frac{(1-\cos\theta)}{2}[15\bar{\lambda}_1(y)\gamma^{\mu}\lambda_2(y)\bar{\lambda}_4(y)\gamma_{\mu}\lambda_3(y)+15\bar{\lambda}_1(y)\gamma^{5}\lambda_2(y)\bar{\lambda}_4(y)\gamma^{5}\lambda_3(y)\nonumber\\
&+&6\bar{\lambda}_1(y)\gamma_{\sigma}\gamma^5\lambda_2(y)\bar{\lambda}_4(y)\gamma^{\sigma}\gamma^5\lambda_3(y)\nonumber\\
&+&\bar{\lambda}_1(y)\gamma^{\nu( \neq \rho)}\gamma^{\rho( \neq \nu)}\gamma^{5}\lambda_2(y)\bar{\lambda}_4(y)\gamma^{5}\gamma_{\rho( \neq \nu)}\gamma_{\nu( \neq \rho)}\lambda_3(y)\nonumber\\
&+&\bar{\lambda}_1(y)\gamma^{5}\gamma^{\rho(\neq \mu)}\gamma^{\mu( \neq \rho)}\lambda_2(y)\bar{\lambda}_4(y)\gamma_{\mu( \neq \rho)}\gamma_{\rho(\neq \mu)}\gamma^{5}\lambda_3(y)\nonumber\\
&+&\bar{\lambda}_1(y)\gamma^{\nu(\neq \mu )}\gamma^{5}\gamma^{\mu(\neq \nu )}\lambda_2(y)\bar{\lambda}_4(y)\gamma_{\mu(\neq \nu )}\gamma^{5}\gamma_{\nu(\neq \mu )}\lambda_3(y)]+ \dots \rbrace\nonumber\\
&\times&\frac{\sin(\frac{\pi \alpha'\tilde{t}}{4})\sin(\frac{\pi \alpha'\tilde{u}}{4})}{\sin(\frac{\pi \alpha'\tilde{s}}{4})}\frac{(\alpha'\tilde{s}/4)^{-\alpha'\tilde{s}/2}(\alpha'\tilde{t}/4)^{\alpha'\tilde{t}/2}(\alpha'\tilde{u}/4)^{\alpha'\tilde{u}/2}}{\tilde{s}^3\frac{(1-\cos\theta)(1+\cos\theta)}{4}} \, . \label{Amplitud-4-dil}
\end{eqnarray}
where we have used the approximation (\ref{gamma-products}) for the product of Gamma functions. At this point, we can use the explicit form of the fermion wave-functions in AdS$_5$, being the four-dimensional spinors \cite{Peskin:1995ev}
\begin{eqnarray}
    u(p)=\begin{pmatrix}
        u_L^s(p) \\ u_R^s(p)
    \end{pmatrix}=\begin{pmatrix}
       \sqrt{p \cdot \sigma} \ \zeta^s \\ \sqrt{p \cdot \bar{\sigma}} \ \zeta^s 
    \end{pmatrix} \, , \label{4d-spinors}
\end{eqnarray}
with $s=1,2$. We must study each term in (\ref{Amplitud-4-dil}). Notice that terms with only one $\gamma^5$ vanish. Since we are interested in the high-energy behavior of the amplitude, i.e. the $s$ power, we can just work out explicitly the first term, while the rest of terms will not change this $s$-behavior in that limit. Then, considering the first term in the integral of equation (\ref{Amplitud-4-dil}) we have
\begin{eqnarray}
&&15\bar{\lambda}_2(y)\gamma^{\mu}\lambda_3(y)\bar{\lambda}_1(y)\gamma_{\mu}\lambda_4(y)=  -15\left(\frac{r}{r_0}\right)^{-\Delta}\frac{\Lambda^2 R^2 (\Delta/2-2)^2}{R^{18}\Lambda^6}  \nonumber \\
&\times & [u_{2,R}^{\dagger}\sigma^{\mu}u_{3,R}u_{1,R}^{\dagger}\sigma_{\mu}u_{4,R}+\frac{M^2R^4}{4r^2 (\Delta_i-3/2)^2}u_{2,L}^{\dagger}\bar{\sigma}^{\mu}u_{3,L}u_{1,R}^{\dagger}\sigma_{\mu}u_{4,R}\nonumber\\
&+&\frac{M^2R^4}{4r^2 (\Delta_i-3/2)^2}u_{2,R}^{\dagger}\sigma^{\mu}u_{3,R}u_{1,L}^{\dagger}\bar{\sigma}_{\mu}u_{4,L}+\frac{M^4R^8}{16r^4(\Delta_i-3/2)^4}u_{2,L}^{\dagger}\bar{\sigma}^{\mu}u_{3,L}u_{1,L}^{\dagger}\bar{\sigma}_{\mu}u_{4,L}] \, , \nonumber\\ \label{first-term}
\end{eqnarray}
where $\Delta=4 \Delta_i$, assuming that there are four identical fermions. In the center-of-mass frame the four-momenta of the four spin-1/2 fermions are
\begin{eqnarray}
    p_1^{\mu}&=&\left(\frac{\sqrt{s}}{2}, \sqrt{\frac{s}{4}-M^2},0,0\right)  \, , \nonumber\\
    p_2^{\mu}&=&\left(\frac{\sqrt{s}}{2}, -\sqrt{\frac{s}{4}-M^2},0,0\right)  \, , \nonumber\\
    p_3^{\mu}&=&\left(-\frac{\sqrt{s}}{2}, \sqrt{\frac{s}{4}-M^2} \cos\theta,\sqrt{\frac{s}{4}-M^2} \sin\theta,0\right)  \, , \nonumber\\
    p_4^{\mu}&=&\left(-\frac{\sqrt{s}}{2},-\sqrt{\frac{s}{4}-M^2} \cos\theta,-\sqrt{\frac{s}{4}-M^2} \sin\theta,0\right)  \, ,
\end{eqnarray}
where $s$ is the four-dimensional Mandelstam variable.
Using this kinematics, the square root factors of the spinors in (\ref{4d-spinors}) become
\begin{eqnarray}
\sqrt{p_1\cdot \sigma}&=&\sqrt{p_2\cdot \bar{\sigma}}=\left(\frac{1}{\sqrt{s}+2M)}\right)^{1/2}\begin{pmatrix}
  -\frac{\sqrt{s}}{2}+M & \sqrt{\frac{s}{4}-M^2} \\ \sqrt{\frac{s}{4}-M^2} & -\frac{\sqrt{s}}{2}+M  
 \end{pmatrix}   \, ,  \nonumber \\
\sqrt{p_2\cdot \sigma} &=&\sqrt{p_1\cdot \bar{\sigma}}=\left(\frac{1}{\sqrt{s}+2M)}\right)^{1/2}\begin{pmatrix}
  -\frac{\sqrt{s}}{2}+M & -\sqrt{\frac{s}{4}-M^2} \\ -\sqrt{\frac{s}{4}-M^2} & -\frac{\sqrt{s}}{2}+M 
 \end{pmatrix}      \, , \nonumber \\
\sqrt{p_3 \cdot \sigma}&=&\sqrt{p_4 \cdot \bar{\sigma}}=\left(\frac{1}{-\sqrt{s}+2M}\right)^{1/2}\begin{pmatrix}
        \frac{\sqrt{s}}{2}+M & \sqrt{\frac{s}{4}-M^2}e^{-i\theta} \\ \sqrt{\frac{s}{4}-M^2}e^{i\theta} & \frac{\sqrt{s}}{2}+M   
    \end{pmatrix}      \, , \nonumber \\
    \sqrt{p_4 \cdot \sigma} &=&\sqrt{p_3 \cdot \bar{\sigma}}=\left(\frac{1}{-\sqrt{s}+2M}\right)^{1/2}\begin{pmatrix}
        \frac{\sqrt{s}}{2}+M & -\sqrt{\frac{s}{4}-M^2}e^{-i\theta} \\ -\sqrt{\frac{s}{4}-M^2}e^{i\theta} & \frac{\sqrt{s}}{2}+M   
    \end{pmatrix}    \, .
\end{eqnarray}
For simplicity let us consider the two-dimensional Weyl spinor
\begin{eqnarray}
\zeta=\begin{pmatrix}
    1 \\ 0
\end{pmatrix} \, .
\end{eqnarray}
Then, the term (\ref{first-term}) is given by
\begin{eqnarray}
&&15\bar{\lambda}_2(y)\gamma^{\mu}\lambda_3(y)\bar{\lambda}_1(y)\gamma_{\mu}\lambda_4(y)= -30\left(\frac{r}{r_0}\right)^{-\Delta}\frac{\Lambda^2R^2(\Delta/2-2)^2}{R^{18}\Lambda^6} \nonumber \\
&\times & (s-4M^2)\left(e^{2i\theta}+\frac{1}{8}e^{i\theta}+1\right) \left(1+\frac{M^2R^4}{2r^2(\Delta_i-3/2)^2}+\frac{M^4R^8}{16r^4(\Delta_i-3/2)^4}\right) \, . \nonumber\\ \label{first-term-1}
\end{eqnarray}
From the analysis of the term (\ref{first-term-1}) its contribution to the four spin-1/2 fermions scattering amplitude is 
\begin{eqnarray}
\mathcal{A}^{4 d}_{\text{4 fermions}}(p) &\approx & \frac{60 e^2 \pi g_s^2 \alpha'^4 (1-\cos\theta)}{R^{14-2\Delta}\Lambda^{4-\Delta} \sin^2\theta}(\Delta/2-2)^2  (s-4 M^2) \nonumber \\ 
&\times &\left(e^{2 i \theta}+\frac{1}{8}e^{i\theta}+1\right)  \int_{r_0}^{\infty} dr \ \left(r^{3-\Delta}+r^{1-\Delta}\frac{M^2R^4}{2(\Delta/4-3/2)^2}\right. \nonumber \\
&& \left. + r^{-1-\Delta}\frac{M^4R^8}{16(\Delta/4-3/2)^4}\right) e^{-2\beta_{\tilde{s}\tilde{t}\tilde{u}}} \, , 
\end{eqnarray}
where as in the case of glueballs we neglected the oscillating factor in the Mandelstam variables since it will not modify the power behavior of $s$ in the amplitude. The function $\beta_{\tilde{s}\tilde{t}\tilde{u}}$ has been defined in (\ref{beta}). We may change the radial variable as in the case of the glueball scattering  $u=r/r_0$, and also take into account the relation $\tilde{s}=(R^2/r^2) s$. After integration we obtain
\begin{eqnarray}
& ~ &\mathcal{A}^{4 d}_{\text{4 fermions}}(p)
\approx  \frac{60 e^2\pi \alpha' (g_s N_c)^{1/2}(1-\cos\theta)}{N_c^2 \sin^2\theta}(\Delta/2-2)^2 \left(e^{2i\theta}+\frac{1}{8}e^{i\theta}+1\right) \times  \nonumber\\
&& \left( 2^{\Delta-5} (s^{2-\tau/2}-4 M^2 s^{1-\tau/2}) \left(-\frac{\Lambda^2 R^2}{\alpha' f(\theta)}\right)^{\Delta/2-2} \Bigg[\Gamma\left(\frac{\Delta-4}{2}\right)-\Gamma\left(\frac{\Delta-4}{2}, - \frac{\alpha' s f(\theta)}{4\Lambda^2 R^2}\right)\Bigg] \right. \nonumber\\
&& \left. +\frac{2^{\Delta-3}(s^{1 - \tau/2}-4M^2s^{- \tau/2} )}{2 (\Delta/4-3/2)^2}  \left(-\frac{\Lambda^2 R^2}{\alpha' f(\theta)}\right)^{\Delta/2-1} \Bigg[\Gamma
   \left(\frac{\Delta-2}{2}\right)-\Gamma
   \left(\frac{\Delta-2}{2},-\frac{\alpha' s f(\theta))}{4
   \Lambda^2 R^2}\right)\Bigg] \right. \nonumber\\
&&\left. +\frac{2^{\Delta-1}(s^{-\tau/2}-4M^2s^{-1-\tau/2}) }{16 (\Delta/4-3/2)^4} \left(-\frac{\Lambda^2 R^2}{\alpha' f(\theta)
   )}\right)^{\Delta/2} \Bigg[\Gamma
   \left(\frac{\Delta}{2}\right)-\Gamma \left(\frac{\Delta}{2},-\frac{\alpha' s f(\theta)}{4
   \Lambda^2 R^2}\right)\Bigg] \right) \label{A-4-fermions}
\end{eqnarray}
where we have used that $\tau=\sum_{i=1}^4\tau_i=\sum_{i=1}^4(\Delta_i-s_i)=\Delta-4\times 1/2=\Delta-2\rightarrow \Delta=\tau+2$. Let us emphasize that in this kinematic domain there is a leading contribution to $\mathcal{A}^{4 d}_{\text{4 fermions}}(p)$ which behaves like $(\alpha's)^{2-\tau/2}$, and also a sub-leading one which goes like $(\alpha's)^{1-\tau/2}$. It is easy to explicitly check this behavior of the three terms within the parenthesis of equation (\ref{A-4-fermions}) considering a physical parametric range, for instance setting $\Lambda \approx M \approx 1$ GeV and for large values of $s$. This implies that the total cross section for this exclusive process involving four spin-1/2 fermions at large $s$ and fixed scattering angle given by
\begin{eqnarray}
\sigma_{\text{Total}} = \frac{1}{s} \ \text{Im}\mathcal{A}^{4d}_{4 \ {\text{fermions}}}(p) \, ,
\end{eqnarray}
contains a leading contribution given by $(s/\Lambda^2)^{2-\tau/2}$,  where the total twist $\tau$ indicates the minimum total number of elementary fields which carry a finite fraction of the total momentum of the exclusive process.     

~

Now, let us consider the Regge limit, i.e. $\tilde{s} \gg |\tilde{t}|$. The Regge limit of the ten-dimensional scattering amplitude of four dilatinos has been obtained in our work \cite{MPS}, and reads
\begin{eqnarray}
&&\mathcal{A}_{\text{string}}^{\text{4 dilatinos Regge}}(\tilde{1},\tilde{2},\tilde{3},\tilde{4}) = -4\pi g_s^2 \alpha'^7 \ K_{\text{string}}^{\text{4 dilatinos Regge}}(\tilde{1},\tilde{2},\tilde{3},\tilde{4}) \nonumber \\
&\times & \frac{\sin[\frac{\pi\alpha'}{4}(\tilde{s}+\tilde{t})]}{\sin(\frac{\pi\alpha'\tilde{s}}{4})} \left(\frac{\alpha'\tilde{s}}{4}\right)^{-2+\frac{\alpha'\tilde{t}}{2}} e^{2-\frac{\alpha'\tilde{t}}{2}}\frac{\Gamma(-\alpha'\tilde{t}/4)}{\Gamma(1+\alpha'\tilde{t}/4)} \nonumber \\ 
&=& 2\pi \ 4^{-\frac{\alpha'\tilde{t}}{2}} g_s^2 \alpha'^4 (\alpha'\tilde{s})^{1+\frac{\alpha'\tilde{t}}{2}}[5(\bar{\lambda}_2\Gamma^{0}\lambda_3 \bar{\lambda}_1\Gamma_{0}\lambda_4+\bar{\lambda}_2\Gamma^{1}\lambda_3 \bar{\lambda}_1\Gamma_{1}\lambda_4) \nonumber \\
&+&\bar{\lambda}_2\Gamma^{0}\lambda_3\bar{\lambda}_1\Gamma_{1}\lambda_4+\bar{\lambda}_2\Gamma^{1}\lambda_3\bar{\lambda}_1\Gamma_{0}\lambda_4+6\bar{\lambda}_2\Gamma^{i_{>1}}\lambda_3\bar{\lambda}_1\Gamma_{{i_{>1}}}\lambda_4] \nonumber \\
&\times & \frac{\sin[\frac{\pi\alpha'}{4}(\tilde{s}+\tilde{t})]}{\sin(\frac{\pi\alpha'\tilde{s}}{4})} e^{2-\frac{\alpha'\tilde{t}}{2}}\frac{\Gamma(-\alpha'\tilde{t}/4)}{\Gamma(1+\alpha'\tilde{t}/4)} \, . \label{A-Regge-10-dimensions-dilatinos}
\end{eqnarray}

In order to make easier the calculation and without lost of generality we focus on the first term. Thus, we have
\begin{eqnarray}
&&\mathcal{A}_{\text{4 fermions}}^{4d \ {\text{Regge}}}(\tilde{1},\tilde{2},\tilde{3},\tilde{4})\approx \nonumber \\
&& 10 e^2 \pi  g_s^2 \alpha'^4\int_{r_0}^{\infty} dr (r^3 R^2)(\alpha'\tilde{s})^{1+\frac{\alpha'\tilde{t}}{2}}\lambda^{\dagger}_2(y)\lambda_3(y)\lambda^{\dagger}_1(y) \lambda_4(y)\frac{\Gamma(-\alpha'\tilde{t}/4)}{\Gamma(1+\alpha'\tilde{t}/4)} \nonumber \\
&=& \frac{10e^2\pi g_s^2(\Delta/2-2)^2 \alpha'^4(s-4M^2)(e^{i\theta}-1)^2}{R^{14-2\Delta}\Lambda^{4-\Delta}}  \Big(\int_{r_0}^{\infty} dr \frac{r^{5-\Delta}}{\alpha'R^2|t|}(\alpha'R^2s/r^2)^{1+\frac{\alpha'R^2t/r^2}{2}} \nonumber \\
&+&\frac{M^2R^4}{2(\Delta/4-3/2)^2}\int_{r_0}^{\infty} dr \frac{r^{3-\Delta}}{\alpha'R^2|t|}(\alpha'R^2s/r^2)^{1+\frac{\alpha'R^2t/r^2}{2}} \nonumber \\
&+&\frac{M^4R^8}{16(\Delta/4-3/2)^4}\int_{r_0}^{\infty} dr \frac{r^{1-\Delta}}{\alpha'R^2|t|}(\alpha'R^2s/r^2)^{1+\frac{\alpha'R^2t/r^2}{2}}\Big) \, .
\end{eqnarray}
Notice that $\bar{\lambda}_i(y)=\lambda_i^\dagger(y) \Gamma^0$.
Next, we must use the saddle-point approximation on the following three integrals in the scattering amplitude $\mathcal{A}_{\text{4 fermions}}^{4d \ {\text{Regge}}}(\tilde{1},\tilde{2},\tilde{3},\tilde{4})$ as follows,
\begin{eqnarray}
&&\mathcal{A}_{\text{4 fermions}}^{4d \ {\text{Regge}}}(\tilde{1},\tilde{2},\tilde{3},\tilde{4})\approx \nonumber \\
&&
\frac{10e^2\pi g_s^2(\Delta/2-2)^2 \alpha'^4(s-4M^2)(e^{i\theta}-1)^2}{R^{14-2\Delta}\Lambda^{4-\Delta}}\nonumber\\
&\times&\Bigg(\int_{r_0}^{\infty} dr \exp\Big[\log\Big(\frac{r^{5-\Delta}}{\alpha'R^2|t|}(\alpha'R^2s/r^2)^{1+\frac{\alpha'R^2t/r^2}{2}}\Big)\Big]\nonumber\\
&+&\frac{M^2R^4}{2(\Delta/4-3/2)^2}\int_{r_0}^{\infty} dr \exp\Big[\log\Big(\frac{r^{3-\Delta}}{\alpha'R^2|t|}(\alpha'R^2s/r^2)^{1+\frac{\alpha'R^2t/r^2}{2}}\Big)\Big]\nonumber\\
&+&\frac{M^4R^8}{16(\Delta/4-3/2)^4}\int_{r_0}^{\infty} dr \exp\Big[\log\Big(\frac{r^{1-\Delta}}{\alpha'R^2|t|}(\alpha'R^2s/r^2)^{1+\frac{\alpha'R^2t/r^2}{2}}\Big)\Big]\Bigg) \, .
\end{eqnarray}
Defining the following functions
\begin{eqnarray}
    f_1(r)&=&\log\Big(\frac{r^{5-\Delta}}{\alpha'R^2|t|}(\alpha'R^2s/r^2)^{1+\frac{\alpha'R^2t/r^2}{2}}\Big) \, ,\\
    f_2(r)&=&\log\Big(\frac{r^{3-\Delta}}{\alpha'R^2|t|}(\alpha'R^2s/r^2)^{1+\frac{\alpha'R^2t/r^2}{2}}\Big) \, ,\\
    f_3(r)&=&\log\Big(\frac{r^{1-\Delta}}{\alpha'R^2|t|}(\alpha'R^2s/r^2)^{1+\frac{\alpha'R^2t/r^2}{2}}\Big) \, ,
\end{eqnarray}
the saddle-point conditions require 
\begin{eqnarray}
    \frac{df_1(r_1^*)}{dr}=\frac{df_2(r_2^*)}{dr}=\frac{f_3(r_3^*)}{dr}=0 \, ,
\end{eqnarray}
whose solutions are
\begin{eqnarray}
    r_1^*&\approx&R\sqrt{\frac{\alpha'|t|\log(s/|t|)}{\Delta-3}}\, , \\
      r_2^*&\approx&R\sqrt{\frac{\alpha'|t|\log(s/|t|)}{\Delta-1}} \, , \\
        r_3^*&\approx&R\sqrt{\frac{\alpha'|t|\log(s/|t|)}{\Delta+1}} \, ,
\end{eqnarray}
being the leading $r$ for each integral 
\begin{eqnarray}
 r_1'&=&\text{min}\left(\Lambda R^2,\Lambda R^2\sqrt{\frac{\hat{\alpha'}|t|\log\left(\frac{s}{|t|}\right)}{\Delta-3}}\right) \, , \\
 r_2'&=&\text{min}\left(\Lambda R^2,\Lambda R^2\sqrt{\frac{\hat{\alpha'}|t|\log\left(\frac{s}{|t|}\right)}{\Delta-1}}\right) \, , \\
  r_3'&=&\text{min}\left(\Lambda R^2,\Lambda R^2\sqrt{\frac{\hat{\alpha'}|t|\log\left(\frac{s}{|t|}\right)}{\Delta+1}}\right) \, ,
\end{eqnarray}
respectively. We may obtain the Regge behavior of this scattering amplitude when the following set of conditions is satisfied
 $\sqrt{\frac{\hat{\alpha'}|t|\log\left(\frac{s}{|t|}\right)}{\Delta-3}}>1, \ \sqrt{\frac{\hat{\alpha'}|t|\log\left(\frac{s}{|t|}\right)}{\Delta-1}}>1, \ \sqrt{\frac{\hat{\alpha'}|t|\log\left(\frac{s}{|t|}\right)}{\Delta+1}}>1$. Thus,
\begin{eqnarray}
&&\mathcal{A}_{\text{4 fermions}}^{4d \ {\text{Regge}}}(\tilde{1},\tilde{2},\tilde{3},\tilde{4}) \approx 
 \frac{5 g_s^2 e^2 \pi^2 \Lambda R^2}{16 \hat{\alpha'}|t|}  \alpha'^4 (\Delta-4)^2  \left(-1+e^{i \theta}\right)^2 \nonumber \\
&\times & \Lambda^{\Delta+1} R^{2 (\Delta-4)} \left(\Lambda R^2\right)^{-\Delta}\left(\sqrt{4 \pi g_s N_c}(\hat{\alpha'}s)^{2+\frac{\hat{\alpha'} t}{2}}-4(\hat{\alpha'}s)^{1+\frac{\hat{\alpha'} t}{2}}\right)\nonumber\\
&\times&   \Bigg(\frac{256 }{(\Delta-6)^4 \sqrt{\Bigg|-3 \hat{\alpha'} |t| \log \left(\hat{\alpha'} s\right)-5
   \hat{\alpha'}|t|+(\Delta+1)\Bigg|}} \nonumber \\
&+&\frac{128}{(\Delta-6)^2 \sqrt{\Bigg|-3 \hat{\alpha'} |t|\log \left(\hat{\alpha'} s\right)-5\hat{\alpha'}
   |t|+(\Delta-1)\Bigg| }}\nonumber\\
   &+&\frac{16}{\sqrt{\Bigg|-3 \hat{\alpha'} |t|
   \log \left(\hat{\alpha'} s\right)-5\hat{\alpha'}|t|+(\Delta-3)\Bigg|}}\Bigg) \, ,
\end{eqnarray}
where we have set $\Lambda\sim M$. If we consider that $\hat{\alpha'}|t|\sim 1$, we obtain
\begin{eqnarray}
 &&\mathcal{A}_{\text{4 fermions}}^{4d \ {\text{Regge}}}(\tilde{1},\tilde{2},\tilde{3},\tilde{4})
  \approx \frac{5 e^2 \pi ^{1/2}\alpha'\Lambda^2\sqrt{4 \pi g_s N_c} }{128 N_c^2} (\Delta-4)^2  \left(-1+e^{i \theta}\right)^2 \nonumber \\
   &\times & \left(\sqrt{g_s N_c}(\hat{\alpha'}s)^{2+\frac{\hat{\alpha'} t}{2}}-4(\hat{\alpha'}s)^{1+\frac{\hat{\alpha'} t}{2}}\right)  \Bigg[\frac{256 }{(\Delta-6)^4 \sqrt{\Bigg|\Delta-4-3 \log \left(\hat{\alpha'} s\right) \Bigg|}} \nonumber \\
&+&\frac{128}{(\Delta-6)^2 \sqrt{\Bigg|\Delta-6-3 \log \left(\hat{\alpha'} s\right)\Bigg| }} + \frac{16}{\sqrt{\Bigg|\Delta-8-3 
   \log \left(\hat{\alpha'} s\right)\Bigg|}}\Bigg] \, . \label{A-Regge-4-dilatinos}
\end{eqnarray}
It is important to understand the meaning of the leading and the sub-leading contributions to $\mathcal{A}_{\text{4 fermions}}^{4d \ {\text{Regge}}}(\tilde{1},\tilde{2},\tilde{3},\tilde{4})$ in equation (\ref{A-Regge-4-dilatinos}). Firstly, let us emphasize that $\mathcal{A}_{\text{string}}^{\text{4 dilatinos Regge}}(\tilde{1},\tilde{2},\tilde{3},\tilde{4})$ in (\ref{A-Regge-10-dimensions-dilatinos}) is the result of the Regge limit on the full ten-dimensional scattering amplitude of 4 dilatinos, i.e. it contains all the corresponding terms directly derived from superstring theory \cite{MPS}. The leading contribution is proportional to $(\hat{\alpha'}s)^{2+\frac{\hat{\alpha'} t}{2}}$. Thus, we may think that there is a Regge trajectory  
$j(t)=2+\frac{\hat{\alpha'}t}{2}=2+\frac{t}{2\sqrt{\lambda}\Lambda^2}$, which can be understood as the Reggeization of a graviton propagating between the two spin-1/2 fermions in four dimensions. Also notice that this leading contribution is enhanced by a factor given by the squared root of the 't Hooft coupling, which is larger than one. On the other hand, there is also a sub-leading contribution, with a Regge trajectory $j(t)=1+\frac{\hat{\alpha'}t}{2}=1+\frac{t}{2\sqrt{\lambda}\Lambda^2}$, which is obtained from the Reggeization of the vector field $B_\mu^1$ which is a linear combination of the gravi-photon $h_{\mu\alpha}$ and the Ramond-Ramond four-form field $A_{\mu\alpha\beta\delta}$ as explained in the introduction \cite{Kovensky:2018xxa,Jorrin:2022lua}.

%
%
\section{Discussion and conclusions}
%
%

We have investigated the high-energy behavior of scattering amplitudes for exclusive two-to-two scattering processes at fixed angle. We have also studied the corresponding Regge limit. In section 2 we have carried out a detailed calculation involving four glueballs following reference \cite{Polchinski:2001tt}. In section 3, where the main results of the present work have been developed, we have obtained the behavior of the scattering amplitude in both limits involving four spin-1/2 fermions.

The differential cross-section for exclusive scattering at large $s$ and fixed $t/s$ is given by \cite{Brodsky:1973kr,Brodsky:1974vy}.
\begin{eqnarray}
\frac{d \sigma}{dt} = \frac{1}{16 \pi s^2} |\mathcal{A}_{2 \rightarrow 2}^{4d}(s, t)|^2 \propto s^{2-\tau} f(t/s) \, , \label{Brodsky-Farrar}
\end{eqnarray}
where $f(t/s)$ is a function of the scattering angle. In the case of two-to-two scattering of glueballs from the amplitude (\ref{A4-glueballs-final}) we obtain 
\begin{eqnarray}
\frac{d \sigma}{dt} \propto \left(\frac{s}{\Lambda^2}\right)^{2-\Delta} f_{\text{4 glueballs}}(t/s) \, ,
\end{eqnarray}
where $\tau=\Delta=\sum_{i=1}^4 \Delta_i$ since the spin of the dual operators is zero, while $f_{\text{4 glueballs}}(t/s)$ can be extracted from equation (\ref{A4-glueballs-final}). One may study the effective supergravity action as the low-energy limit of type IIB superstring theory. In this case the relevant effective action is \cite{Polchinski:2002jw}
\begin{eqnarray}
S_{\text{effective}}^{\text{dilatons}} = \int d^{10}x \sqrt{g_{10}} A^m v^j \partial_m\phi \ \partial_j\phi \, , \label{Seff-dilaton}
\end{eqnarray}
where the dilaton $\phi$ is a charge eigenstate with charge $\cal{Q}$, while the Killing vector $v^j$ satisfies the equation
\begin{eqnarray}
v^j \partial_j Y^s(\Omega) = i {\cal {Q}} Y^s(\Omega) \, .
\end{eqnarray}
The action (\ref{Seff-dilaton}) can be derived from the type IIB supergravity action considering a non-diagonal metric perturbation of the form $\delta g_{m j}=A_m(y, r) \ v_j(\Omega)$.
Therefore, the interaction between two dilatons and the gravi-photon $A_m$ cannot change the conformal dimension of the spin-zero glueball operators ${\cal{O}}_k^{(8)}(x)$, given by linear combinations of ${\cal {N}}=4$ SYM operators of the form $\text{Tr}(F_+^2 X^k)$, with $\Delta=k + 4$ (being $k \geq 0$). Thus, for exclusive two-to-two glueball scattering processes there are the following two possibilities, namely: $\Delta_1=\Delta_2$ and $\Delta_3=\Delta_4$, or $\Delta_1=\Delta_4$ and $\Delta_2=\Delta_3$, while for the elastic scattering we have the obvious condition $\Delta_1=\Delta_4$ and $\Delta_2=\Delta_3$.
In the first case a single gravi-photon is exchanged in the $s$-channel. In the other two cases, in the dual model there is an exchange of a single gravi-photon in the $t$-channel of type IIB supergravity. This adds some restrictions to the selection rules coming from the angular integrals on $S^5$ of the four scalar spherical harmonics discussed in section 2. In addition, in this context we can also make contact with the small angle scattering amplitude where the exponent of the factor $(\hat{\alpha'}s)^{2+\frac{\hat{\alpha'} t}{2}}$ has been interpreted as a Regge trajectory $j(t)=2+\frac{\hat{\alpha'}t}{2}=2+\frac{t}{2\sqrt{\lambda}\Lambda^2}$, obtained from the Reggeization of a graviton propagating between the two glueballs in the $t$-channel. 

\vspace{0.5cm}

On the other hand, it is very interesting to compare our results for spin-1/2 fermions to the scaling laws at large transverse momentum obtained by Brodsky and Farrar \cite{Brodsky:1973kr,Brodsky:1974vy} for proton-proton to proton-proton scattering. For large $s \gg M^2$ the leading contribution to the scattering amplitude from equation (\ref{A-4-fermions}) gives
\begin{eqnarray}
\frac{d \sigma}{dt} \propto \left(\frac{s}{\Lambda^2}\right)^{2-\tau} f_{\text{4 fermions}}(t/s) \, ,
\end{eqnarray}
where $\tau=\sum_{i=1}^4 \tau_i$. Also $f_{\text{4 fermions}}(t/s)$ can be extracted from equation (\ref{A-4-fermions}). Let us consider the minimal twist operators of  ${\cal {N}}=4$ SYM theory of the form ${\cal{O}}_k^{(6)}(x)$ that can be written as linear combinations of operators of the form $\text{Tr}(F_+ \lambda_{{\cal {N}}=4} X^k)$, being $\lambda_{{\cal {N}}=4}$ the gauginos of the ${\cal {N}}=4$ gauge supermultiplet. The minimal twist of each individual operator is $\tau_i=3$ (i.e. for $k=0$), the total twist is $\tau=4 \times \tau_i=12$, therefore, 
\begin{eqnarray}
\frac{d \sigma}{dt}\bigg|_{\text{minimal twist}} \propto \left(\frac{s}{\Lambda^2}\right)^{-10} f_{\text{4 fermions}}(t/s) \, .
\end{eqnarray}
This result gives exactly the same scaling behavior for the differential cross section as the one obtained in references  \cite{Brodsky:1973kr,Brodsky:1974vy} for proton-proton to proton-proton scattering. Although our result and the one of these references give the same scaling behavior we should emphasize that the ${\cal {N}}=4$ SYM theory fermionic operator $\text{Tr}(F_+ \lambda_{{\cal {N}}=4})$ obviously does not represent a proton. Recall that the twist counts the minimal number of internal constituents participating in the scattering process, i.e. three valence quarks in a proton. Another very important distinction is the fact that in QCD the concept of partons is  inherently associated with perturbation theory, while the gauge/string description studied here uses perturbative string theory and therefore ${\cal {N}}=4$ SYM theory must be strongly coupled.  Notwithstanding, our result seems to add evidence toward a description of the dynamics of protons in terms of string theory scattering amplitudes in the gauge/string theory duality framework. In particular, as explained in the introduction the BPST Pomeron and the Holographic $A_4$ Pomeron (also for ${\cal {N}}=4$ SYM theory with an infrared cut-off) both fit DIS experimental data of the proton with remarkable accuracy. A possible explanation for this success is that somehow the dual string theory description captures certain universal properties for confining non-Abelian gauge theories in four dimensions.

One may study the effective supergravity action involving dilatinos as the low-energy limit of type IIB superstring theory. In references \cite{Jorrin:2020cil,Jorrin:2020kzq} the dilatino part of this action has been derived from direct dimensional reduction on $S^5$ of the type IIB supergravity action. This leads to
\begin{eqnarray}
S^{\text{dilatinos}}_{\text{effective}}&=& K \int dz \ d^4x \sqrt{-g_{AdS_5}} \times \nonumber \\
&& \ \left( i \frac{{\cal{Q}}}{3}
\bar{\lambda}^-_{k} \tilde{\gamma}^a B^1_a\lambda^-_{k} +i \ \frac{b^{-, -}_{1 k j}}{12} \bar{\lambda}^-_{j} F^{ab}  \tilde{\Sigma}_{ab} \lambda^-_k + i  \frac{b^{+, -}_{1 k j}}{12} \bar{\lambda}^+_{j} F^{ab} \tilde{\Sigma}_{ab} \lambda^-_k\right) \, , \label{five-dimensional-action} 
\end{eqnarray}
being
\begin{eqnarray}
b^{\pm,-}_{1 k j}&=&\left(1+ 2 \left(k\mp j+\frac{5}{2}\mp\frac{5}{2}\right) \right) \int d \Omega_5  (\Theta^{\pm}_{j})^{\dag} \tau_{\alpha}v^{\alpha}\Theta_{k} +4{\cal{Q}}  \int   d \Omega_5 (\Theta^{\pm}_{j})^{\dag}  \Theta^-_{k}  \, ,
\end{eqnarray}
where there is a normalization constant $K$. Also, there is a vector field, $B_a^1$, which is the massless Maxwell-Einstein field in AdS$_5$ defined as a certain linear combination of off-diagonal fluctuations of the metric tensor and vector fluctuations of the Ramond-Ramond four-field potential,
\begin{equation}
B_a^1(x) \equiv A_a^1(x) - 16 \Phi_a^1(x) \, , \label{Bfield}
\end{equation}
where $A_a^1(x)$ is a metric fluctuation given by
\begin{equation}
h_{a \alpha} = \sum_{I_5} A_a^{I_5}(x) \, Y^{I_5}_\alpha(y) \, ,
\end{equation}
while the field $\Phi_a^1(x)$ comes from the mode expansion of the Ramond-Ramond field,
\begin{equation}
a_{a \alpha\beta\gamma} = \sum_{I_5} \Phi_a^{I_5}(x)\, \epsilon_{\alpha\beta\gamma\delta\epsilon} \, \nabla^\delta Y^{I_5 \epsilon}(y) \, .
\end{equation}
The effective fermionic action is also written in terms of the two-form field strength, where the covariant derivative is indicated
\begin{eqnarray}
F_{ab}=\nabla_a B_b^1-\nabla_b B_a^1 \, ,
\end{eqnarray}
and $\tilde{\Sigma}_{ab}=\frac{1}{4}(\tilde{\gamma}_a \tilde{\gamma}_b-\tilde{\gamma}_b \tilde{\gamma}_a)$. To complete the definitions, notice that $I_5$ is a compact notation for the set of numbers $(l_5,l_4,l_3,l_2,l_1)$ labelling $Y^{I_5 \epsilon}(y)$, i.e. the vector spherical harmonics on $S^5$. In the present case, $B_a^1(x)$ are the 15 Yang-Mills fields of {\bf {15}} irreducible vector representation $SU(4) \sim SO(6)$, while the vector spherical harmonics are Killing vectors of $S^5$. For the right-handed dilatinos on AdS$_5 \times S^5$ we have 

\begin{equation}
    \lambda(y, \Omega)=\left({\begin{array}{c}
   0\\
   \lambda'(y, \Omega) \\
  \end{array} }\right) \, , 
\end{equation}
where
\begin{eqnarray}
\lambda'(y, \Omega)=\sum_k \left(\lambda^+_k(y) \Theta_k^+(\Omega) + \lambda_k^{-}(y)
\Theta_k^-(\Omega)\right) \, , \label{spinor-expansion}
\end{eqnarray}
while $\Theta_k^+(\Omega)$ and  $\Theta_k^-(\Omega)$ are spinor spherical harmonics on $S^5$. There are two towers of masses associated with the irreducible representations {\bf 4$^*$}, {\bf 20$^*$}, {\bf 60$^*$}, $\cdots$ ($-$), or {\bf 4}, {\bf 20}, {\bf 60}, $\cdots$ ($+$) of the $SO(6)$ isometry group of $S^5$, which are labelled as $\lambda^{\pm}_k$  on the AdS$_5$.

Then, the first term in the action (\ref{five-dimensional-action}) corresponds to the minimal coupling. Basically, it is
related to the leading term $s^{2-\tau/2}$ in the scattering amplitude ${\cal {A}}^{4d}_{\text{4 fermions}}$, and it is associated with the exchange of a gravi-photon. The other terms in (\ref{five-dimensional-action}) lead to the exchange of a vector field between two dilatinos in type IIB supergravity on AdS$_5 \times S^5$, and are related to   the sub-leading term $s^{1-\tau/2}$ in scattering amplitude ${\cal {A}}^{4d}_{\text{4 fermions}}$. The Reggeization of these fields, namely: the gravi-photon and the vector field, lead to the Regge trajectories $j(t)=2+\frac{\hat{\alpha}' t}{2}$ and $j(t)=1+\frac{\hat{\alpha}' t}{2}$, respectively, indicated in the last paragraph of section 3. Also, in addition to the selection rules which may be derived from the integrals of four spinor spherical harmonics, there are restrictions from the different interaction terms in the action (\ref{five-dimensional-action}). For instance, the first term preserves the conformal dimension of the ${\cal {N}}=4$ SYM operators, while the others allow for certain mixing.

~

%
\centerline{\large{\bf Acknowledgments}}
%

~

We thank David Jorrin for useful discussions.
The work of L.M., M.P. and M.S. has been supported in part by the Consejo Nacional de Investigaciones Cient\'{\i}fi\-cas y T\'ecnicas of Argentina (CONICET). This work has been partially supported by the CONICET Grants PIP-UE B\'usqueda de nueva f\'{\i}sica and PICT-E 2018-0300 (BCIE).

\newpage

\appendix

\section{Spherical harmonics on $S^5$} \label{Appendix-A}

In this appendix we review the construction of scalar spherical harmonics on $S^5$, and construct explicitly some scalar spherical harmonics which are useful for section 2. The relation to the Kaluza-Klein states obtained from spontaneous compactification of type IIB supergravity on AdS$_5 \times S^5$ was derived in the pioneering article \cite{Kim:1985ez}. Also, it is very useful the review article \cite{DHoker:2002nbb} to give the explicit relation between ${\cal {N}}=4$ SYM theory operators and the Kaluza-Klein states obtained in \cite{Kim:1985ez}.

Let us construct the scalar spherical harmonics on $S^5$. We have to consider the irreducible representations of $SU(4)$ with Dynkin integers $[0, k, 0]$, with $k \ge 0$. There is a relation between the second Dynkin integer $k$ and $\Delta$ given by $\Delta=k+4$. The dimension of the $[0, k, 0]$ representation is obtained by associating a Young diagram with $k$ columns of two boxes each. The dimension of this irreducible representation is given by the ratio of two Young tableaux, which leads to \cite{vanNieuwenhuizen:2012zk,vanNieuwenhuizen:2019lbe}
\begin{equation}
d_{\text{scalar}}(5, k) =  \left[ \left( \begin{matrix} 5+k \\ 5  \end{matrix}  \right) - \left( \begin{matrix} 5+k-2 \\ 5 \end{matrix}  \right) \right] \,  .\label{dimension-scalar-S5}
\end{equation}
Also, the spherical harmonics on $S^5$ are labelled by $(l_1,l_2,l_3,l_4,l_5)$, where $l_i \ge 0$ are integers. They satisfy $l_5 \ge l_4 \ge l_3 \ge l_2 \ge l_1 \ge 0$ and $l_5 \equiv k$.

The scalar spherical harmonics on $S^{n-1}$ have be constructed in \cite{Higuchi:1986wu}
\begin{equation}
Y_{(l_{n-1}, \dots, l_1)}^{(s)}(\theta_1, \dots, \theta_{n-1})= (-1)^{l_1} \ \frac{1}{\sqrt{2\pi}} \ e^{i l_1 \theta_1} \ \prod_{j=2}^{n-1}  \  _j{\bar{P}}^{l_{j-1}}_{l_j}(\theta_j) \,  , \label{YscalarSn}
\end{equation}
where $\theta_j$'s are the $n-1$ angles on $S^{n-1}$. In this construction it has been used the Legendre functions defined as
\begin{equation}
_j{\bar{P}}^l_L(\theta)=\sqrt{\frac{2L+j-1}{2}\frac{(L+l+j-2)!}{(L-l)!}} \ (\sin\theta)^{\frac{2-j}{2}} \ P_{\left(L+\frac{j-2}{2}\right)}^{-\left(l+\frac{j-2}{2}\right)}(\cos\theta) \, ,
\end{equation}
in terms of the associated Legendre polynomials $P_{\left(L+\frac{j-2}{2}\right)}^{-\left(l+\frac{j-2}{2}\right)}(\cos\theta)$. The scalar spherical harmonics satisfy
\begin{equation}
\Box_{S^{n-1}} Y_{(l_{n-1}, \dots, l_1)}^{(s)}(\theta_1, \dots, \theta_{n-1})=- l_{l_n-1} \ (l_{l_n-1} + n - 2) \ Y_{(l_{n-1}, \dots, l_1)}^{(s)}(\theta_1, \dots, \theta_{n-1}) \, .
\end{equation}

Since we are interested in the $SU(N_c)$ ${\cal{N}}=4$ SYM theory, which is dual to type IIB superstring theory on AdS$_5 \times S^5$, in this case $n-1=5$.

Next, let us consider some examples. For the $[0,0,0]$ irreducible representation of $SU(4)$, where $k=0$, its dimension is one, and the spherical harmonic is given in equation (\ref{Y00000}). Setting $k=1$ the irreducible  representation is $[0,1,0]$ of dimension 6, which can be seen as the ratio of the values of the following Young tableaux:
\begin{equation}
\Yvcentermath1
\Yboxdim{12pt}
{\scriptsize\young(4,3)} \,\,\,\,\, \text{and}  \,\,\,\,\,
{\scriptsize\young(2,1)} \ . 
\end{equation}
We have explicitly obtained the 6 scalar spherical harmonics, 
\begin{eqnarray}
Y_{(1,0,0,0,0)}^{(s)} &=&  \frac{\sqrt{6}}{\pi^{3/2}} \ \cos\theta_5 \ , \\
Y_{(1,1,0,0,0)}^{(s)} &=&  \frac{\sqrt{6}}{\pi^{3/2}} \ \cos\theta_4 \ \sin\theta_5 \ , \\
Y_{(1,1,1,0,0)}^{(s)} &=&  \frac{2\sqrt{6}}{\pi^{3/2}} \ \frac{ \cos\theta_3 \ \sin\theta_4 \ \sin^2\left(\frac{\theta_5}{2}\right) \ (1+\cos\theta_5)}{\sin\theta_5} \, , \\
Y_{(1,1,1,1,0)}^{(s)} &=&  \frac{2\sqrt{6}}{\pi^{3/2}} \ \frac{ \cos\theta_2 \ \sin\theta_3 \ \sin\theta_4 \ \sin^2\left(\frac{\theta_5}{2}\right) \ (1+\cos\theta_5)}{\sin\theta_5} \, , \\
Y_{(1,1,1,1,1)}^{(s)} &=&  \frac{2\sqrt{3}}{\pi^{3/2}} \ e^{-i \theta_1} \ \frac{ \sin\theta_2 \ \sin\theta_3 \ \sin\theta_4 \ \sin^2\left(\frac{\theta_5}{2}\right) \ (1+\cos\theta_5)}{\sin\theta_5} \, , \\
Y_{(1,1,1,1,-1)}^{(s)} &=&  \frac{2\sqrt{3}}{\pi^{3/2}} \ e^{i \theta_1} \ \frac{ \sin\theta_2 \ \sin\theta_3 \ \sin\theta_4 \ \sin^2\left(\frac{\theta_5}{2}\right) \ (1+\cos\theta_5)}{\sin\theta_5} \, .
\end{eqnarray}
We have checked their orthonormality and calculated the integral of four scalar spherical harmonics on $S^5$.

For $k=2$, the representation is $[0,2,0]$, and there are 20 orthogonal scalar spherical harmonics on $S^5$, which can be calculated from the ratio of the values of the following Young tableaux:
\begin{equation}
\Yvcentermath1
\Yboxdim{12pt}
{\scriptsize\young(45,34)} \,\,\,\,\, \text{and}  \,\,\,\,\,
{\scriptsize\young(32,21)} \ . 
\end{equation}
The corresponding 20 scalar spherical harmonics can be easily obtained from equation (\ref{YscalarSn}), but we do not to write them here. For construction of spinor spherical harmonics on $S^5$ we refer the reader to references \cite{Higuchi:1986wu,Camporesi:1995fb,Jorrin:2020cil,Jorrin:2020kzq}.

\pagebreak

\end{document}